\documentclass[%
reprint,
superscriptaddress,
 amsmath,amssymb,
 aps
]{revtex4-2}
 
\renewcommand{\figurename}{\textbf{Fig.}}
\renewcommand*{\thefigure}{\textbf{\arabic{figure}}}
\renewcommand{\tablename}{\textbf{Table}}
\renewcommand*{\thetable}{\textbf{\arabic{table}}}
\usepackage{soul}
\usepackage{tikz}
\usetikzlibrary{tikzmark}
\usepackage{multirow}
\usepackage{graphicx}
\usepackage{dcolumn}
\usepackage{bm} 
\usepackage{float}
\usepackage[colorlinks,
            urlcolor=blue,
            linkcolor=blue,
           anchorcolor=blue,
            citecolor=blue]{hyperref}
\usepackage{mfirstuc}
\usepackage{natbib}
\MFUnocap{for}
\MFUnocap{a}
\MFUnocap{of}
\MFUnocap{an}
\MFUnocap{by}
\MFUnocap{to}
\MFUnocap{on}
\MFUnocap{into}
\MFUnocap{with}
\MFUnocap{from}
\MFUnocap{via}
\MFUnocap{and}
\MFUnocap{beyond}
\MFUnocap{in}
\MFUnocap{between}
\MFUnocap{to}
\MFUnocap{as}
\MFUnocap{over}
\MFUnocap{the}
\usepackage{lettrine}
\RequirePackage[explicit]{titlesec}
\renewcommand{\thesubsection}{\Alph{subsection}}
\titleformat{\section}
  {\large\sffamily\bfseries}
  {\thesection.}
  {0em}
  {\capitalisewords{#1}}
  []
\titleformat{name=\section,numberless}
  {\normalsize\sffamily\bfseries}
  {}
  {0em}
  {{#1}}
  [] 
\titleformat{\subsection}
  {\normalsize\rmfamily\bfseries}
  {\thesubsection.}
  {0em}
  {#1}
  []
\titlespacing*{\section}{0pt}{4.5ex plus 2ex minus .2ex}{.0ex plus .0ex}
\titlespacing*{\subsection}{0pt}{3.2ex plus 2ex minus .2ex}{.0ex plus .0ex}

\begin{document}


\preprint{APS/123-QED}

\title{{Directional Dipole Dice Enabled by Anisotropic Chirality}}

\author{Yuqiong Cheng}
\affiliation{Department of Physics, City University of Hong Kong, Tat Chee Avenue, Kowloon, Hong Kong, China}
\author{Kayode Adedotun Oyesina}
\author{Bo Xue}
\affiliation{Department of Electrical Engineering, City University of Hong Kong, Tat Chee Avenue, Kowloon, Hong Kong, China}
\author{Dangyuan Lei}
\affiliation{Department of Materials Science and Engineering, City University of Hong Kong, Tat Chee Avenue, Kowloon, Hong Kong, China}
\author{Alex M. H. Wong}\email{alex.mh.wong@cityu.edu.hk}
\affiliation{Department of Electrical Engineering, City University of Hong Kong, Tat Chee Avenue, Kowloon, Hong Kong, China}
\affiliation{State Key Laboratory of Terahertz and Millimeter Waves, City University of Hong Kong, Tat Chee Avenue, Kowloon, Hong Kong, China}
\author{Shubo Wang}\email{shubwang@cityu.edu.hk}
\affiliation{Department of Physics, City University of Hong Kong, Tat Chee Avenue, Kowloon, Hong Kong, China}
\affiliation{City University of Hong Kong Shenzhen Research Institute, Shenzhen, Guangdong 518057, China}
\date{\today}
            
\begin{abstract}
Directional radiation and scattering play an essential role in light manipulation for various applications in integrated nanophotonics, antenna and metasurface designs, quantum optics, etc. The most elemental system with this property is the class of directional dipoles, including the circular dipole, Huygens dipole, and Janus dipole. A unified realization of all three dipole types and a mechanism to freely switch among them are previously unreported, yet highly desirable for developing compact and multifunctional directional sources. Here, we theoretically and experimentally demonstrate that the synergy of chirality and anisotropy can give rise to all three directional dipoles in one structure at the same frequency under linearly polarized plane wave excitations. This mechanism enables a simple helix particle to serve as a directional dipole dice (DDD), achieving selective manipulation of optical directionality via different “faces” of the particle. We employ three “faces” of the DDD to realize face-multiplexed routing of guided waves in three orthogonal directions with the directionality determined by spin, power flow, and reactive power, respectively. This construction of the complete directionality space can enable the unprecedented high-dimensional control of both near-field and far-field directionality with broad applications in photonic integrated circuits, quantum information processing, and subwavelength-resolution imaging.
\end{abstract}

\maketitle


\section*{\textcolor[RGB]{200,36,38}{INTRODUCTION}}
\noindent
Controlling the propagation direction of light to achieve directional radiation or scattering is a key objective of light manipulations, with important applications in almost every aspect of photonics and plasmonics \cite{lin2013polarization,wang2014lateral,petersen2014chiral,lodahl2017chiral,xie2020metasurface,chen2020chiral}. Far-field directional radiation can be realized by applying the design principle of high-directivity antennas \cite{kosako2010directional,novotny2011antennas} or engineering the interference of electric and magnetic multipoles to satisfy Kerker conditions as in Huygens antennas \cite{fu2013directional,yao2016controlling,ziolkowski2017using}. Near-field directional routing can be achieved by manipulating the local polarization or symmetries of confined fields, leading to the discovery of directional dipole sources \cite{rodriguez2013near,picardi2018janus,zhong2021toggling} and directional meta-sources \cite{long2020designing}. There are three types of elemental directional dipoles: circular dipole, Huygens dipole, and Janus dipole. The circular dipole (i.e., circularly polarized electric/magnetic dipole) can excite unidirectionally propagating guided waves via spin-momentum locking \cite{o2014spin,le2015nanophotonic,bliokh2015spin,van2016universal,picardi2017unidirectional,shi2021transverse,shi2022intrinsic}, with fascinating applications in topological photonics and non-Hermitian physics \cite{luo2017spin,wang2019arbitrary,yang2020spin} as well as in designing novel nanophotonic devices \cite{sayrin2015nanophotonic,espinosa2017chip,chen2021long}. The Huygens dipole can give rise to directional power flow in both the near and far fields \cite{jin2010metamaterial,geffrin2012magnetic,coenen2014directional,nechayev2019huygens}, which can be employed to achieve vanished backscattering \cite{person2013demonstration}, cloaking \cite{pfeiffer2013metamaterial}, perfect reflection and refraction \cite{wong2018perfect,chen2018huygens}, and near-field optical microscopy \cite{arango2022cloaked}. The Janus dipole has side-dependent directional properties derived from the reactive power, exhibiting complete near-field coupling or noncoupling to waveguides
\cite{picardi2018janus,picardi2019experimental,chen2020directional,wu2020intrinsic}.

The directional dipoles are usually realized by using different optical structures because of their different physical mechanisms. The circular dipole, composed of a pair of orthogonal electric/magnetic dipoles $\pm \pi / 2$ out of phase, can be realized in plasmonic nanospheres under the excitation of circularly polarized light \cite{petersen2014chiral,o2014spin}. The Huygens dipole and Janus dipole, composed of orthogonal electric and magnetic dipoles in phase and $\pm \pi / 2$ out of phase, respectively, can be realized by tailoring high-index dielectric nanospheres or nanocylinders supporting electric and magnetic Mie resonances \cite{wozniak2015selective,picardi2019experimental,picardi2022integrated}. To facilitate the development of high-dimensional and multifunctional directional sources for applications in integrated photonics and quantum optics, it is highly desired to realize all the three directional dipoles and freely switch among them in $\textit{one structure}$ and at the $\textit{same frequency}$. However, this seems to be an unattainable goal considering that they require different compositions of dipoles with different relative phases and amplitudes.

Here, we demonstrate a general physical mechanism for the unified realization of all three directional dipoles and the construction of a complete directionality space. We show that the synergy of chirality and anisotropy can give rise to circular dipole, Huygens dipole, and Janus dipole in the same structure and at the same frequency, under the excitation of a linearly polarized plane wave. Such anisotropic chirality can be found in many structures, such as a simple helix particle made of metal. The magnetoelectric coupling of the helix particle enables the excitation of electric and magnetic dipoles at the same frequency, and the anisotropy enables the excitation of different dipoles in different directions. The relative phase differences of the excited electric and magnetic dipoles naturally satisfy the requirements of the directional dipoles. Such a helix chiral particle gives rise to the three directional dipoles in three orthogonal directions, which enable selective manipulation of all types of optical directionality pointing in orthogonal directions via different faces of the helix, corresponding to a directional dipole dice (DDD). We show that the switch of different directional dipoles can be easily achieved by tuning the propagation and polarization directions of the incident plane wave. Using a mode expansion theory, we uncover that the emergence of the directional dipoles is attributed to the selective excitation of the plasmonic resonance modes in the helix. To characterize the unique properties of the DDD, we employ multiple directional-dipole faces to achieve multiplexed unidirectional excitation of guided waves and experimentally demonstrate the phenomena at microwave frequencies.

\section*{\textcolor[RGB]{200,36,38}{RESULTS}}

\subsection*{Mechanism and realization}
\noindent
The directional dipoles correspond to a combination of electric and/or magnetic dipoles whose magnitudes and phases satisfy certain conditions. The circular electric and magnetic dipoles can be defined as
\begin{equation}
    \mathbf{D}_{\text {cir }}^{\mathrm{e}}=\left(p \hat{\mathbf{e}}_{i}, \pm \mathrm{i} p \hat{\mathbf{e}}_{j}\right), \mathbf{D}_{\text {cir }}^{\mathrm{m}}=\left(m \hat{\mathbf{e}}_{i}, \pm \mathrm{i} m \hat{\mathbf{e}}_{j}\right), 
     \label{eq:1}
\end{equation}
where $p$ and $m$ are the magnitudes of the electric and magnetic dipole components, respectively. Here and in what follows, $\hat{\mathbf{e}}_{i}$ and $\hat{\mathbf{e}}_{j}$ denote the unit axis vectors in Cartesian coordinate system with $i,j=x,y,z$ and $i \neq j$. The directionality of the circular dipoles is given by their spin $\mathbf{S}=\operatorname{Im}\left[\left(\mathbf{D}_{\text {cir }}^{\mathrm{e}, \mathrm{m}}\right)^{*} \times \mathbf{D}_{\text {cir }}^{\mathrm{e}, \mathrm{m}}\right]$. The Huygens dipole can be defined as \cite{kerker1983electromagnetic} 
\begin{align}
    \mathbf{D}_{\text {Huy }}=\left(p \hat{\mathbf{e}}_{i}, \pm m \hat{\mathbf{e}}_{j}\right) \text { with } p=\frac{m}{c}. 
    \label{eq:2}
\end{align}
Here, $c$ is the speed of light in vacuum. The directionality of the Huygens dipole is given by the time-averaged power flow (i.e., the real part of the Poynting vector): $\operatorname{Re}[\mathbf{P}]=\frac{1}{2} \operatorname{Re}\left[\mathbf{E} \times \mathbf{H}^{*}\right]$. The Janus dipole can be defined as \cite{picardi2018janus}
\begin{align}
    \mathbf{D}_{\mathrm{Jan}}=\left(p \hat{\mathbf{e}}_{i}, \pm \mathrm{i} m \hat{\mathbf{e}}_{j}\right) \text { with } p=\frac{m}{c}. 
    \label{eq:3}
\end{align}
The directionality of the Janus dipole is given by the reactive power (i.e., the imaginary part of the Poynting vector): $\operatorname{Im}[\mathbf{P}]=\frac{1}{2} \operatorname{Im}\left[\mathbf{E} \times \mathbf{H}^{*}\right]$ \cite{picardi2018janus}. Therefore, we can assign the directions of the spin $\mathbf{S}$, the power flow $\operatorname{Re}[\mathbf{P}]$, and the reactive power $\operatorname{Im}[\mathbf{P}]$ to be the directions of the circular dipole, the Huygens dipole, and the Janus dipole, respectively.

The above directional dipoles can be realized by using passive structures, such as subwavelength particles under the excitation of electromagnetic waves. For an isotropic achiral particle, the induced electric and magnetic dipoles can be expressed as $\mathbf{p}=\alpha_{\mathrm{ee}}\mathbf{E}$ and $ \mathbf{m}=\alpha_{\mathrm{mm}}\mathbf{B}$, respectively, where $\mathbf{E}$ $(\mathbf{B})$ is the incident electric (magnetic) field and $\alpha_{\mathrm{ee}}$ ($\alpha_{\mathrm{mm}}$) is the electric (magnetic) polarizability. For a circularly polarized incident plane wave $\mathbf{E}=(\hat{\mathbf{e}}_{x} \pm \mathrm{i} \hat{\mathbf{e}}_{y}) E_{0} e^{\mathrm{i} k z}$, this particle can give rise to circular electric dipole $\mathbf{D}_{\mathrm{cir}}^{\mathrm{e}}=\left(\hat{\mathbf{e}}_{x}, \pm \mathrm{i} \hat{\mathbf{e}}_{y}\right) \alpha_{\mathrm{ee}} E_{0}$ and circular magnetic dipole $\mathbf{D}_{\mathrm{cir}}^{\mathrm{m}}=\left(\hat{\mathbf{e}}_{x}, \pm \mathrm{i} \hat{\mathbf{e}}_{y}\right) \alpha_{\mathrm{mm}} E_{0}/c$. However, it is impossible to simultaneously achieve the Huygens and Janus dipoles, which require different relative phases between the electric and magnetic dipoles (as in Eqs.~(\ref{eq:2}) and (\ref{eq:3})).

For an isotropic chiral particle, the induced dipoles can be expressed as 
\begin{align}
\left[\begin{array}{c}
\mathbf{p} \\
\mathbf{m}
\end{array}\right]=\left[\begin{array}{cc}
\alpha_{\mathrm{ee}} & \mathrm{i} \alpha_{\mathrm{em}} \\
-\mathrm{i} \alpha_{\mathrm{em}} & \alpha_{\mathrm{mm}}
\end{array}\right]\left[\begin{array}{l}
\mathbf{E} \\
\mathbf{B}
\end{array}\right],
\label{eq:4}
\end{align}
where $\alpha_{\mathrm{em}}$ denotes the magnetoelectric polarizability derived from the chirality of the particle. Under the excitation of a linearly polarized incident plane wave $\mathbf{E}=\hat{\mathbf{e}}_{x} E_{0} e^{\mathrm{i} k z}$, the circular electric dipole can be induced as $\mathbf{D}_{\mathrm{cir}}^{\mathrm{e}}=\left(\alpha_{\mathrm{ee}} \hat{\mathbf{e}}_{x}, \mathrm{i} \frac{\alpha_{\mathrm{em}}}{c} \hat{\mathbf{e}}_{y}\right) E_{0}$ when $\alpha_{\mathrm{ee}}=\pm \alpha_{\mathrm{em}} / c$, and circular magnetic dipole can be induced as $\mathbf{D}_{\mathrm{cir}}^{\mathrm{m}}=\left(-\mathrm{i} \alpha_{\mathrm{em}} \hat{\mathbf{e}}_{x}, \frac{\alpha_{\mathrm{mm}}}{c} \hat{\mathbf{e}}_{y}\right) E_{0}$ when $\alpha_{\mathrm{em}}=\pm \alpha_{\mathrm{mm}} / c$. In addition, the Huygens dipole can be induced as $\mathbf{D}_{\text {Huy }}=\left(\alpha_{\mathrm{ee}} \hat{\mathbf{e}}_{x}, \frac{\alpha_{\mathrm{mm}}}{c} \hat{\mathbf{e}}_{y}\right) E_{0}$ when $\alpha_{\mathrm{ee}}=\pm \alpha_{\mathrm{mm}} / c^{2}$, and it can also be given by $\mathbf{D}_{\text {Huy }}=\left(\frac{1}{c} \hat{\mathbf{e}}_{y},-\hat{\mathbf{e}}_{x}\right) \mathrm{i} \alpha_{\mathrm{em}} E_{0}$ via the magnetoelectric polarizability $\alpha_{\mathrm{em}}$. Finally, the Janus dipole can be induced as $\mathbf{D}_{\mathrm{Jan}}=\left(\alpha_{\mathrm{ee}} \hat{\mathbf{e}}_{x}, \frac{\alpha_{\mathrm{mm}}}{c} \hat{\mathbf{e}}_{y}\right) E_{0}$ when $\alpha_{\mathrm{ee}}=\pm \mathrm{i} \alpha_{\mathrm{mm}} / c^{2}$. Therefore, all the three directional dipoles can be simultaneously realized in an isotropic chiral particle satisfying the following conditions: 
\begin{align}
    \alpha_{\mathrm{ee}}=\pm \frac{\mathrm{i} \alpha_{\mathrm{mm}}}{c^{2}}=\pm \frac{\alpha_{\mathrm{em}}}{c} \text { or } \alpha_{\mathrm{ee}}=\pm \frac{\mathrm{i} \alpha_{\mathrm{mm}}}{c^{2}}=\pm \frac{\mathrm{i} \alpha_{\mathrm{em}}}{c}.
    \label{eq:5}
\end{align}
The above mechanisms are summarized in Table \ref{tab:table1}, where the arrows indicate the combinations of polarizabilities that give rise to the desired directional dipoles.
\bgroup
\begin{table}[h]
\caption{\label{tab:table1}%
\textbf{Chirality-enabled directional dipoles}}
\centering
\def\arraystretch{1.2}
\setlength\tabcolsep{8pt}
\begin{tabular}{lrcc}
 \hline 
 \hline 
 & \multicolumn{2}{c}{Polarizability} & Condition \\ 
 \hline
 
\multirow{2}{*}{Circular dipole} & $\alpha_{\text{ee}}$ \tikzmark{c11}  & \tikzmark{c12} i$\alpha_{\text{em}}$ & $\alpha_{\text{ee}}=\pm\alpha_{\text{em}}/c$ \\
 & $-\text{i}\alpha_{\text{em}}$ \tikzmark{c21} & \tikzmark{c22} $\alpha_{\text{mm}}$ & $\alpha_{\text{em}}=\pm\alpha_{\text{mm}}/c$ \\
\multirow{2}{*}{Huygens dipole} & $\alpha_{\text{ee}}$ \tikzmark{h11} & \tikzmark{h12} i$\alpha_{\text{em}}$ & \multirow{2}{*}{N.A.} \\
 & $-\text{i}\alpha_{\text{em}}$ \tikzmark{h21} & \tikzmark{h22} $\alpha_{\text{mm}}$ &  \\
\multirow{2}{*}{Janus dipole} & $\alpha_{\text{ee}}$ \tikzmark{j11} & \tikzmark{j12} i$\alpha_{\text{em}}$ & \multirow{2}{*}{$\alpha_{\text{ee}}=\pm \text{i}\alpha_{\text{mm}}/c^2$} \\
 & $-\text{i}\alpha_{\text{em}}$ \tikzmark{j21} & \tikzmark{j22} $\alpha_{\text{mm}}$ & \\
 \hline 
 \hline 
\end{tabular}

\begin{tikzpicture}[overlay, remember picture, shorten >=.5pt, shorten <=.5pt, transform canvas={yshift=.25\baselineskip}]
    \draw [<->,line width=0.5pt] ({pic cs:c11}) -- ({pic cs:c12});
    \draw [<->,line width=0.5pt] ({pic cs:c21}) -- ({pic cs:c22});
    \draw [<->,line width=0.5pt] ({pic cs:h21}) -- ({pic cs:h12});
    \draw [<->,line width=0.5pt] ({pic cs:j11}) -- ({pic cs:j22});
  \end{tikzpicture}

\end{table}
\bgroup

Although the above $\mathbf{D}_{\text {cir }}^{\mathrm{e}, \mathrm{m}}, \mathbf{D}_{\mathrm{Huy}}, \text { and } \mathbf{D}_{\mathrm{Jan}}$ can be realized simultaneously, their directions overlap. In practical applications, it is desired to have the directional dipoles point in orthogonal directions to construct a complete directionality space, which can enable the multiplexed control of directionality and the realization of combined directional sources. This can be achieved by combining chirality with anisotropy. For an anisotropic chiral particle, the scalar polarizabilities in Eq. (\ref{eq:4}) are replaced by the polarizability tensors$\overleftrightarrow{\alpha}_{\mathrm{ee}}$, $\overleftrightarrow{\alpha}_{\mathrm{mm}}$ and $\overleftrightarrow{\alpha}_{\mathrm{em}}$. The anisotropy provides additional degrees of freedom to realize the circular dipole, Huygens dipole, and Janus dipole in orthogonal directions. We consider, for example, a helix particle with the center axis in $y$ direction. The magnetoelectric polarizability tensor $\overleftrightarrow{\alpha}_{\mathrm{em}}$ is dominated by the component $\alpha_{\mathrm{em}}^{y y}$. Under the tilted incidence of a linearly polarized plane wave $\mathbf{E}=\left(E_{y} \hat{\mathbf{e}}_{y}+E_{z} \hat{\mathbf{e}}_{z}\right) e^{\mathrm{i} k_{y} y+\mathrm{i} k_{z} z}$, the circular electric dipole can be induced in $\pm y$ direction as $\mathbf{D}_{\text {cir }}^{\mathrm{e}}=\left(\alpha_{\mathrm{ee}}^{x y} E_{y} \hat{\mathbf{e}}_{x}, \alpha_{\mathrm{ee}}^{z z} E_{z} \hat{\mathbf{e}}_{z}\right)$  when $\operatorname{Arg}\left(\alpha_{\mathrm{ee}}^{x y}\right)-\operatorname{Arg}\left(\alpha_{\mathrm{ee}}^{z z}\right)=\pm \pi / 2$. The Huygens dipole can be induced in $\pm x$ direction as $\mathbf{D}_{\text {Huy }}=\left(\alpha_{\mathrm{ee}}^{z z} E_{z} \hat{\mathbf{e}}_{z},-\mathrm{i} \alpha_{\mathrm{em}}^{y y} E_{y} \hat{\mathbf{e}}_{y}\right)$ when $\operatorname{Arg}\left(\alpha_{\mathrm{em}}^{y y}\right)-\operatorname{Arg}\left(\alpha_{\mathrm{ee}}^{z z}\right)=\pm \pi / 2$. The Janus dipole can be induced in  $\pm z$ direction as $\mathbf{D}_{\text {Jan }}=\left(\alpha_{\mathrm{ee}}^{x y} \hat{\mathbf{e}}_{x},-\mathrm{i} \alpha_{\mathrm{em}}^{y y} \hat{\mathbf{e}}_{y}\right) E_{y}$  when  $\alpha_{\mathrm{ee}}^{x y}=\pm \alpha_{\mathrm{em}}^{y y} / c$, or $\mathbf{D}_{\text {Jan }}=\left(\alpha_{\mathrm{ee}}^{y y} E_{y} \hat{\mathbf{e}}_{y}, \alpha_{\mathrm{mm}}^{x x} B_{x} \hat{\mathbf{e}}_{x}\right)$ when  $\operatorname{Arg}\left(\alpha_{\mathrm{ee}}^{y y}\right)-\operatorname{Arg}\left(\alpha_{\mathrm{mm}}^{x x}\right)=\pm \pi / 2$. We note that the relative amplitudes of the field components $E_{y}, E_{z} \text { and } B_{x}$ in the above expressions can be tuned by varying the incident angle. Therefore, the anisotropic chiral particle can simultaneously realize all the three directional dipoles in orthogonal directions at the conditions:
\begin{align}
    \operatorname{Arg}\left(\alpha_{\mathrm{ee}}^{x y}\right)-\operatorname{Arg}\left(\alpha_{\mathrm{ee}}^{z z}\right)=\pm \pi / 2, \alpha_{\mathrm{ee}}^{x y}=\pm \alpha_{\mathrm{em}}^{y y} / c, \label{eq:6}
\end{align}
or
\begin{equation}
    \begin{gathered}
    \operatorname{Arg}\left(\alpha_{\mathrm{ee}}^{x y}\right)-\operatorname{Arg}\left(\alpha_{\mathrm{ee}}^{z z}\right)=\pm \pi / 2,\\ \operatorname{Arg}\left(\alpha_{\mathrm{em}}^{y y}\right)-\operatorname{Arg}\left(\alpha_{\mathrm{ee}}^{z z}\right)=\pm \pi / 2,\\ \operatorname{Arg}\left(\alpha_{\mathrm{ee}}^{y y}\right)-\operatorname{Arg}\left(\alpha_{\mathrm{mm}}^{x x}\right)=\pm \pi / 2.
    \label{eq:7}
    \end{gathered}
\end{equation}
\begin{figure}[tb!]
\centering
\includegraphics{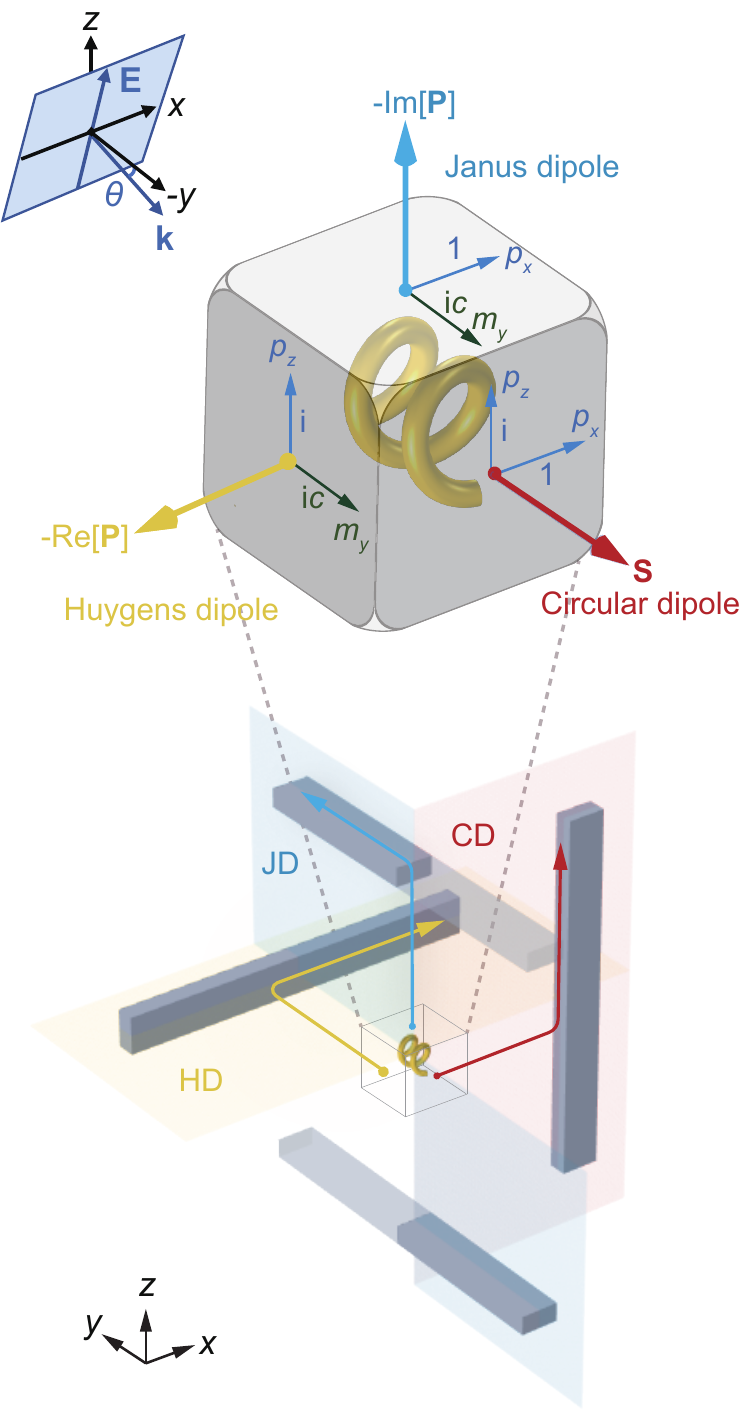}
\caption{\textbf{Schematic of the DDD realized by a metallic helix particle}. The circular dipole, Huygens dipole and Janus dipole feature on three faces of the “dice”. The large arrows denote the directions of the directional dipoles defined by the spin $\mathbf{S}$, the power flow $\operatorname{Re}[\mathbf{P}]$, and the reactive power $\operatorname{Im}[\mathbf{P}]$. The incident linearly polarized plane wave propagates in the $yz$-plane, forming an angle $\theta$ with the $-y$ axis. Bottom inset: the DDD can construct a complete directionality space, achieving face-multiplexed and high-dimensional routing of the guided waves via different dipole faces in different directions.} 
\label{fig1}
\end{figure}

Here, we use the metallic helix particle in Fig. \ref{fig1} to demonstrate the mechanism. Such helices can be fabricated by using low-temperature shadow deposition \cite{mark2013hybrid} and have been extensively studied for its intriguing chiroptical properties \cite{hentschel2017chiral}, such as circular dichroism \cite{gibbs2013plasmonic}, optical forces \cite{wang2014lateral,wo2020optical}, polarization conversion \cite{gansel2009gold}, and vortex beam generation \cite{fang2018broadband}. We assume that the incident plane wave is linearly polarized with the electric field $\mathbf{E}_{\mathrm{inc}}=(-\sin \theta \hat{\mathbf{e}}_{y}+\cos \theta \hat{\mathbf{e}}_{z}) E_{0} e^{\left(-\mathrm{i} k_{0} \cos \theta y-\mathrm{i} k_{0} \sin \theta z\right)}$, where $\theta$ is the incident angle between the wavevector $\mathbf{k} \text { and }-y$ direction and we have neglected the time-harmonic factor $e^{-\mathrm{i} \omega t}$. We will show that the helix can give rise to three dipole components $p_{x}, p_{z} \text { and } m_{y}$ that constitute the three directional dipoles on three faces of the DDD. Importantly, the spin $\mathbf{S}$ of the circular dipole, the net power flow $\operatorname{Re}[\mathbf{P}]$ of the Huygens dipole, and the reactive power $\operatorname{Im}[\mathbf{P}]$ of the Janus dipole point in $-y,+x, \mathrm{and} -z$ directions, respectively, as shown in Fig. \ref{fig1}. The proposed DDD can construct a complete directionality space, which enables selective scattering and coupling from different directional-dipole faces in three orthogonal directions, as shown by the bottom inset of Fig. \ref{fig1}. The DDD is surrounded by three sets of  waveguide channels to illustrate the face-multiplexed and high-dimensional light routing, with each face of the DDD coupled unidirectionally to one waveguide channel. The circular-dipole (CD) face of DDD can excite the guided wave propagating unidirectionally in $+z$ direction denoted by the red arrow. The Huygens-dipole (HD) face of DDD can excite the guided wave propagating unidirectionally in $+x$ direction denoted by the yellow arrow. The Janus-dipole (JD) face of DDD can predominately couple light to the top waveguide. With appropriate loss or termination in the transparent part of the waveguide, the JD can excite the guided wave propagating unidirectionally in $+y$ direction denoted by the blue arrow. In addition, the directionality of each dipole can be flexibly reversed by tuning the incidence. In the JD case, the reversed directionality corresponds to the excitation of the guided wave unidirectionally propagating in $-y$ direction in the bottom waveguide channel. Therefore, the setting in Fig. 1 allows the complete control of near-field directionality in all three orthogonal directions.

Under the incidence of the plane wave, currents and charges will be induced in the helix, and their oscillations give rise to resonances of the helix. We conducted full-wave simulation of the helix and computed its scattering cross section for the incident angle $\theta=90$ degrees. The results are shown in Fig. \ref{fig2}A as the red symbol line. We notice a resonance appears at the frequency of 108 THz. We then apply multipole expansions and decompose the scattering cross section into contributions of multipoles. As seen, the scattering cross section is dominated by the electric dipole ($C_{\mathrm{sca}}^{\mathrm{p}}$) denoted by the red dashed line. The contribution of magnetic dipole ($C_{\mathrm{sca}}^{\mathrm{m}}$), denoted by the blue dashed line, is negligible. The sum $C_{\mathrm{sca}}^{\mathrm{p}}+C_{\mathrm{sca}}^{\mathrm{m}}$ well agrees with the full-wave numerical result, demonstrating the validity of the multipole expansions. The weightings of the electric and magnetic dipoles at the resonance frequency can be tuned by varying the incident angle $\theta$. Figure \ref{fig2}B shows the relative amplitudes of $p_{x}, p_{z} \text { and } m_{y}$ as a function of $\theta$. We notice that $\left|p_{z}\right| /\left|p_{x}\right| \text { and }\left|p_{z}\right| /\left|m_{y} / c\right|$ reduce as $\theta$ increases, which is due to a smaller $z$ component of the incident electric field at a larger $\theta$. Interestingly, $\left|p_{x}\right| /\left|m_{y} / c\right| \approx 1$ for a wide range of the incident angle. Figure \ref{fig2}C shows the relative phases of $p_{x}, p_{z} \text { and } m_{y}$ as a function of $\theta$. As seen, $\operatorname{Arg}\left(p_{z}\right)-\operatorname{Arg}\left(p_{x}\right)=90$ degrees, $\operatorname{Arg}\left(p_{z}\right)-\operatorname{Arg}\left(m_{y}\right)=180$ degrees, and $\operatorname{Arg}\left(p_{x}\right)-\operatorname{Arg}\left(m_{y}\right)=90$ degrees over a broad range of $\theta$. In particular, the induced dipoles satisfy $p_{z} / p_{x}=\mathrm{i}, p_{z} /\left(m_{y} / c\right)=-1, \quad \text {and} \quad p_{x} /\left(m_{y} / c\right)=\mathrm{i}$ simultaneously at $\theta=20$ degrees, corresponding to a circular dipole, a Huygens dipole, and a Janus dipole, respectively. We note that the physical mechanism is robust and can be realized with other geometric parameters or at other frequencies (e.g., microwave frequencies). Another design of the gold helix with a different set of parameters is provided in Supplementary Fig. S2; a realization at microwave frequencies is provided in Fig. \ref{fig8}.

\begin{figure}[tb!]
\centering
\includegraphics{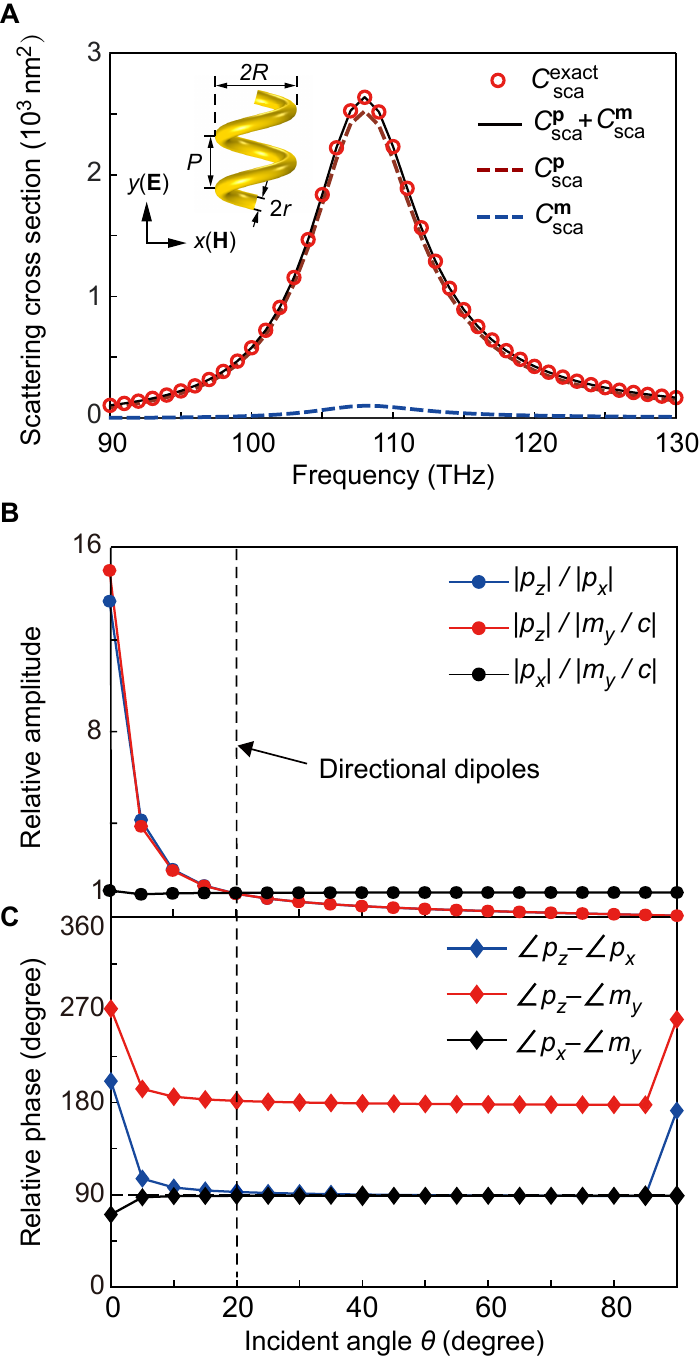}
\caption{\textbf{Electric and magnetic dipole moments induced in the helix}. (\textbf{A}) The scattering cross section of the helix particle and the contributions of the electric and magnetic dipoles. (\textbf{B}, \textbf{C}) The relative amplitudes and phases of the dipole components as a function of the incident angle of the linearly polarized plane wave. The dashed line marks the parameters giving three directional dipoles. } \label{fig2}
\end{figure}

\subsection*{Mode expansion analysis}
The remarkable property of the helix particle can be understood with a mode expansion theory, where we apply three steps to determine its response under external excitations. We first analytically obtain the eigen currents of the helix and use them to construct the Green’s function. Then, we apply the Green’s function to determine the induced currents in the helix under the external excitations. Finally, the induced currents are used to calculate the electric and magnetic dipoles, which constitute the three directional dipoles. The results of this analytical method are compared with the full-wave numerical results to demonstrate its validity and accuracy. 

Since the metal helix can be obtained by twisting a nanorod into helical shape, its eigenmodes correspond to the eigenmodes of the nanorod mapped onto the helical path defined by the helix \cite{hoflich2019resonant}, as long as the helix pitch is large enough so that couplings between helix turns are negligible. The eigenmodes of the nanorod are one-dimensional (1D) standing surface charge waves (i.e., currents) with eigenfrequency $\omega_{n}$ and propagation constant $\gamma_{n}$. These eigen currents are approximately uniform on the cross section of the nanorod since the radius of the nanorod $r \ll \lambda$. The expressions of the eigen currents can be obtained semi-analytically, from which we can determine the eigen currents $\mathbf{J}_n$ of the helix via a mapping. By constructing the Green's function using the eigen currents, we can then analytically determine the response of the helix under arbitrary external excitation, and the induced dipoles can be expanded as: $\mathbf{p}=\sum_{n} B_{n} a_{n} \mathbf{p}_{n}$, $ \mathbf{m}=\sum_{n} B_{n} a_{n} \mathbf{m}_{n}$,
where $\mathbf{p}_{n}=\frac{\mathrm{i}}{\omega} \int \mathbf{J}_{n}(\mathbf{r}) \mathrm{d} V_{\mathrm{p}}$ and $\mathbf{m}_{n}=\frac{1}{2} \int \mathbf{r} \times \mathbf{J}_{n}(\mathbf{r}) \mathrm{d} V_{\mathrm{p}}$ are the dipoles attributed to the eigen current $\mathbf{J}_{n}$, $B_{n}$ is the excited mode amplitude, and $a_{n}(\omega)$ is the excitation-independent expansion coefficient containing holistic resonance characteristics of the helix. The eigen current $\mathbf{J}_n$ with odd value of $n$ gives rise to the dominant dipole components (See Methods)
\begin{equation}
    \begin{gathered}
\left(\mathbf{p}_{n}\right)_{x}=\frac{\mathrm{i} \sigma C R}{\omega}\left(\frac{\sin {\phi_{n}}}{\gamma_{n}-K}+\frac{\sin {\phi_{n}}}{\gamma_{n}+K}\right),\\
\left(\mathbf{p}_{n}\right)_{y}=\frac{\mathrm{i} C P \sin {\phi_{n}}}{\pi \omega \gamma_{n}},\left(\mathbf{p}_{n}\right)_{z}=0,\\
\left(\mathbf{m}_{n}\right)_{y}=\frac{-\sigma C R^{2} \sin {\phi_{n}}}{\gamma_{n}} ,
\label{eq:8}
\end{gathered}
\end{equation}
where $C=2 N \pi^{2} r^{2} / L$, $\phi_{n}=\gamma_{n} L / 2$, $K=2 \pi N / L$, $N$ is the number of turns of the helix, $L$ is arc length of the helix, and $\sigma=+ 1$ ($\sigma=-1$) for the left (right) handed helix. The eigen current $\mathbf{J}_n$ with even value of $n$ gives rise to the dominant dipole components   
\begin{equation}
\begin{gathered}
\left(\mathbf{p}_{n}\right)_{x}=0,\left(\mathbf{p}_{n}\right)_{y}=0,\\
\left(\mathbf{p}_{n}\right)_{z}=\frac{\mathrm{i} C R}{\omega}\left(\frac{\sin \phi_{n}}{\gamma_{n}-K}-\frac{\sin \phi_{n}}{\gamma_{n}+K}\right),\\
\left(\mathbf{m}_{n}\right)_{y}=0 .
\label{eq:9}
\end{gathered}
\end{equation}

\begin{figure*}[htp]
\centering
\includegraphics{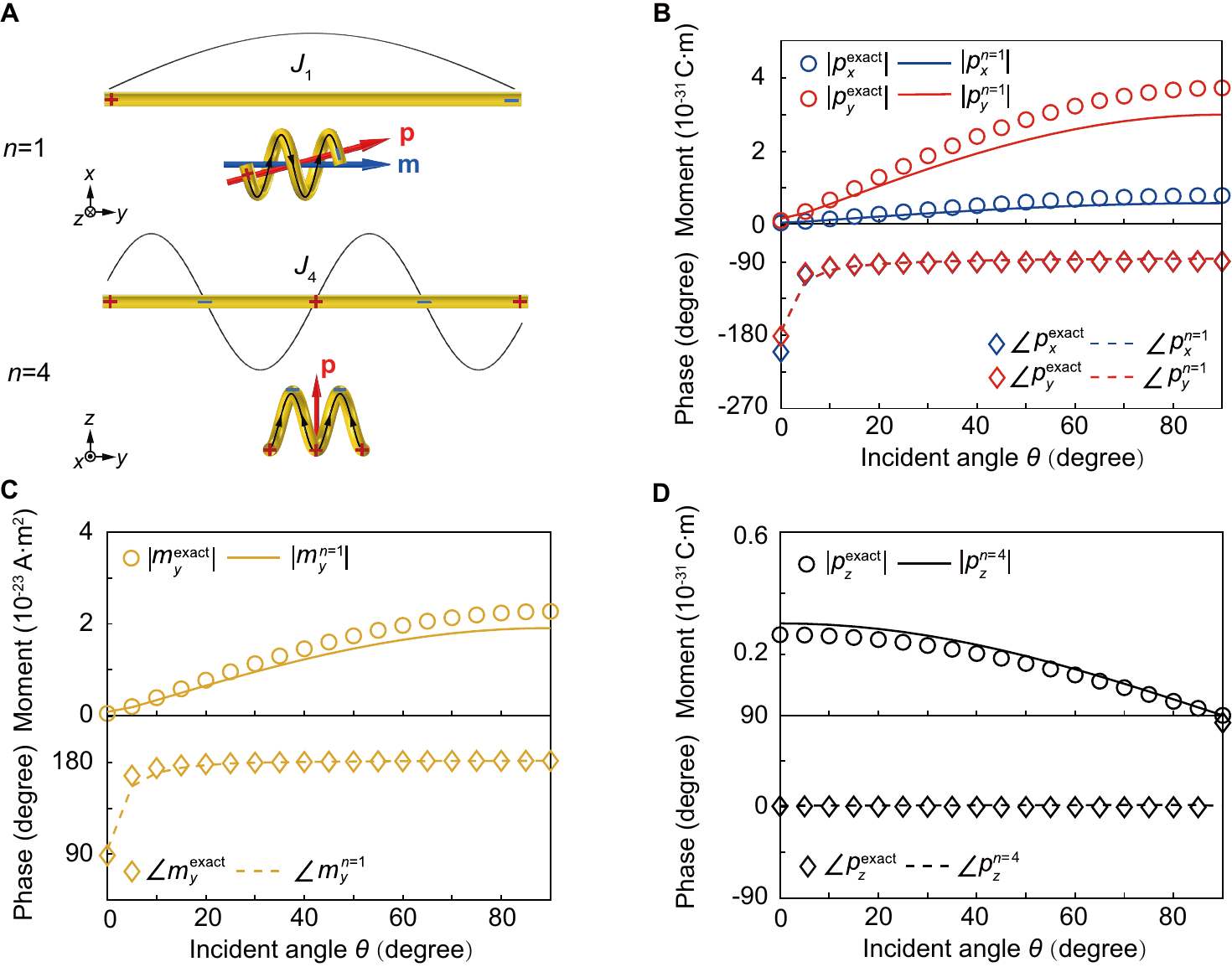}
\caption{\textbf{Mode expansion analysis of the helix}. (\textbf{A}) The charge (denoted by plus and minus symbols) and current (denoted by black arrowed curves) distributions of the first-order and fourth-order eigenmodes of the helix. (\textbf{B}) The amplitudes and phases of the electric dipoles $p_{x}$, $p_{y}$ and (\textbf{C}) the magnetic dipole $m_{y}$ dominating in the first-order eigenmode. (\textbf{D}) The amplitude and phase of the electric dipole $p_{z}$ dominating in the fourth-order eigenmode. The symbols denote the numerical results while the solid lines denote the analytical results of the mode expansions.} \label{fig3}
\end{figure*}

Equations (\ref{eq:8}) and (\ref{eq:9}) indicate that different eigenmodes contribute to different dipole components. The odd-order eigenmodes will generate $p_{x}, p_{y} \text { and } m_{y}$, while the even-order eigenmodes will only generate $p_{z}$. At the eigen frequencies, we have $\gamma_{n} L \approx n \pi$. Thus, $\left(\mathbf{p}_{n}\right)_{z}$ in Eq. (\ref{eq:9}) vanishes for even values of $n$ except for $n=2N$, i.e., $p_{z}$ is mainly contributed by the eigenmode of the order $n=2N$. In addition, the odd-order eigenmodes account for the chirality of the helix because it can generate both electric and magnetic dipoles. The odd- and even-order eigenmodes contribute to electric dipoles in orthogonal directions, which account for the anisotropy of the helix. The synergy of chirality and anisotropy via the two types of eigenmodes can give rise to the desired directional dipoles. In addition, the switching of different directional dipoles can be achieved by selectively exciting the eigenmode, i.e., tuning the mode amplitude $B_{n}$, which depends on the propagation direction and polarization direction of incident field.

We apply the mode expansion theory to analytically determine the induced dipoles in the helix for various incident angles. We find that tilted incidence will predominantly excite the eigenmodes of the orders $n=1,4$.  The charge and current distributions of the two modes are shown in Fig. \ref{fig3}A, where it is evident that they mainly generate dipole components $p_{x}, p_{y}, p_{z} \text { and } m_{y}$. For the $n=1$ eigenmode, the positive and negative charges accumulate at the ends of the helix, giving rise to $p_{x} \text { and } p_{y}$; the current flows in one direction and gives rise to $m_{y}$. Figures \ref{fig3}B and \ref{fig3}C show the comparisons between the analytical (lines) and numerical results (symbols) of $p_{x}, p_{y} \text { and } m_{y}$, which show good agreements. For the $n=4$ eigenmode, the positive and negative charges oscillate along the $z$ direction, generating the dipole component $p_{z}$, as illustrated in Fig. \ref{fig3}A, while the current changes direction periodically in space, leading to vanished magnetic dipole. Figure \ref{fig3}D shows the comparison between the analytical (lines) and numerical (symbols) results of $p_{z}$, which again exhibits good consistency.

The above electric and magnetic dipole components can fulfill the conditions of the three directional dipoles in Eqs. (\ref{eq:1})-(\ref{eq:3}). This can be understood as follows. First, the emergence of both $p_{x}$ and $m_{y}$ in the first eigenmode is attributed to the chirality of the helix, and their relative amplitudes can be tailored by the geometry of the helix to satisfy $\left|p_{x}\right|=\left|m_{y}\right| / c$. In addition, the charge-induced electric dipole $p_{x}$ and the current-induced magnetic dipole $m_{y}$ have an intrinsic phase difference of $ \pi / 2$. Thus, the first eigenmode can give rise to a Janus dipole $\mathbf{D}_{\mathrm{Jan}}=\left(p_{x} \hat{\mathbf{e}}_{x}, m_{y} \hat{\mathbf{e}}_{y}\right)$ with $p_{x} /\left(m_{y} / c\right)=\mathrm{i}$ irrespective of the incident angle. Second, the different values of $p_{x}$ and $p_{z}$ are attributed to the anisotropy of the helix. Their relative amplitude can be tuned by the incident angle of the plane wave because the incident fields $E_{y}$ and $E_{z}$ can excite the first-order and fourth-order eigenmodes, respectively. At an appropriate angle (corresponding to the dashed line in Fig. \ref{fig2}B), one can obtain $\left|p_{x}\right|=\left|p_{z}\right|$. We notice that the first eigenmode is on resonance while the fourth eigenmode is off resonance, which indicates that $p_{x}$ of the first-order eigenmode and $p_{z}$ of the fourth-order eigenmode have a phase difference of $ \pi / 2$. Therefore, the first-order and fourth-order eigenmodes can give rise to a circular electric dipole $\mathbf{D}_{\text {cir }}^{\mathrm{e}}=\left(p_{x} \hat{\mathbf{e}}_{x}, p_{z} \hat{\mathbf{e}}_{z}\right)$  with $p_{z} / p_{x}=\mathrm{i}$. Third, with the combined effect of the chirality and anisotropy of the helix, the first-order and fourth-order eigenmodes can give rise to the Huygens dipole $\mathbf{D}_{\text {Huy }}=\left(p_{z} \hat{\mathbf{e}}_{z}, m_{y} \hat{\mathbf{e}}_{y}\right) \text { with }\left(p_{x} /\left(m_{y} / c\right)\right) \times\left(p_{z} / p_{x}\right)=p_{z} /\left(m_{y} / c\right)=-1$.

The eigenmode analysis can also explain why the phase differences between different dipole components are insensitive to the incident angle $\theta$, corresponding to the results in Fig. \ref{fig2}C. As shown in Fig. \ref{fig3}A, the response of the helix is dominated by the first-order and fourth-order eigenmodes. The first-order eigenmode gives rise to $p_{x}, p_{y} \text { and } m_{y}$. Thus, their phase differences are intrinsic properties of the eigenmode and do not depend on the excitation properties (including the incident angle). This explains the constant phase difference between $p_{x}$ and $m_{y}$, corresponding to the black line in Fig. \ref{fig2}C. In addition, varying the incident angle does not change the relative phase of the excited first-order and fourth-order eigenmodes. This is because their excitations are attributed to the incident electric field only, which is approximately constant over the deep-subwavelength helix and is independent of the incident angle. Therefore, the phase differences between $p_{z}$ of the fourth-order eigenmode and $p_{x}$, $m_{y}$ of the first-order eigenmode (corresponding to the blue and red lines in Fig. \ref{fig2}C) are insensitive to the incident angle. 

\begin{figure}[tp]
\centering
\includegraphics{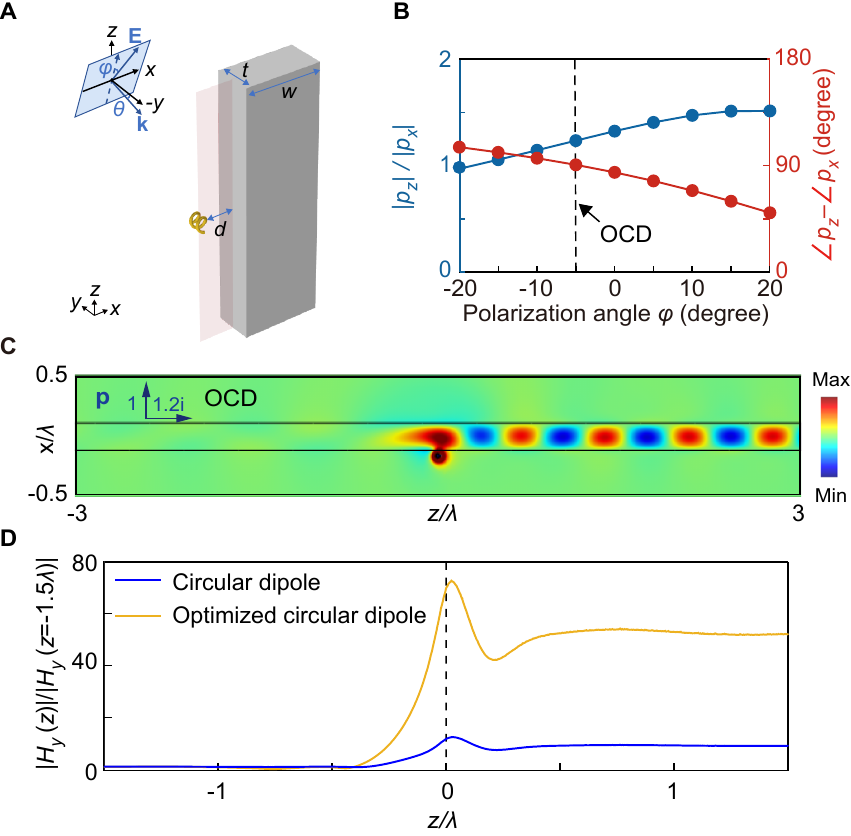}
\caption{\textbf{Directional excitation of guided wave by the circular-dipole face of the helix}. (\textbf{A}) Schematic of the helix-waveguide coupling configuration for demonstrating the directionality of the circular dipole.(\textbf{B}) The relative amplitude and phase of the electric dipoles $p_{z}$ and $p_{x}$ as a function of polarization angle $\varphi$. The incident angle is $\theta=15$ degrees. (\textbf{C}) Unidirectional coupling to the waveguide induced by the optimized circular dipole (OCD) in the helix. (\textbf{D}) The directionality of the ideal circular dipole and the optimized circular dipole in the helix.  } \label{fig4}
\end{figure}

\subsection*{Complete directional excitation of guided waves}
The DDD can be employed to achieve face-multiplexed directional scattering and coupling of electromagnetic waves, corresponding to the scenario in Fig. \ref{fig1}. As a demonstration, we consider the helix located near the surface of a silicon waveguide and under the excitation of an incident plane wave, as shown in Fig. \ref{fig4}A. The coupling between the helix particle and the waveguide can be expressed as \cite{petersen2014chiral}:
\begin{equation}
    \kappa_{\mathrm{pw}} \propto\left|\mathbf{p} \cdot \mathbf{E}^{*}+\mathbf{m} \cdot \mathbf{B}^{*}\right|, 
     \label{eq:10}
\end{equation}
where $\mathbf{E}$ and $\mathbf{B}$ are electric and magnetic fields of the guided mode at the location of the dipoles $\mathbf{p}$ and  $\mathbf{m}$. In the presence of the waveguide, the induced electric and magnetic dipoles are generally different from those of an isolated helix due to the reaction field from the waveguide \cite{wang2016strong}. To achieve high directionality, the directional dipoles can be optimized to satisfy $\left|p_{x} E_{x}^{*}+m_{y} B_{y}^{*}+p_{z} E_{z}^{*}\right|=0$, i.e., the coupling is vanished for a particular guided mode. Following the terminology in Ref. \cite{picardi2018janus}, we call them the optimized directional dipoles to differentiate them from the ideal directional dipoles defined in Eqs. (\ref{eq:1})-(\ref{eq:3}). The optimization can be easily done via tuning the polarization direction of the incident plane wave $\mathbf{E}_{\text {inc }}=(\sin \varphi \hat{\mathbf{e}}_{x}-\sin \theta \cos \varphi \hat{\mathbf{e}}_{y}+\cos \theta \cos \varphi \hat{\mathbf{e}}_{z}) E_{0} e^{-\mathrm{i} k_{0}(\cos \theta y+\sin \theta z)}$, where $\varphi$ is the polarization angle defined as the projection angle of $\mathbf{E}_{\text {inc }}$ on the $zy$-plane, as shown in Fig. \ref{fig4}A. Thus, we can tune the polarization direction by varying the angle $\varphi$. 

We first demonstrate the directional excitation of guided wave with the circular-dipole face of the DDD. The silicon waveguide supports a fundamental $\mathrm{TE}$ guided mode at the dipole resonance frequency of the helix (i.e., $108$ THz). As shown in Fig. 4A, the helix is located $d=80$ nm near the waveguide surface parallel to $yz$-plane, and its axis is in the $y$-direction to switch on the directionality of electric circular dipole $\mathbf{D}_{\mathrm{cir}}^{\mathrm{e}}=\left(p_{x} \hat{\mathbf{e}}_{x}, p_{z} \hat{\mathbf{e}}_{z}\right)$. By adjusting the incident angle $\theta$ and the polarization angle $\varphi$, the helix can realize an optimized circular dipole (i.e., elliptical dipole) with $p_{z} / p_{x}=-E_{x}^{*} / E_{z}^{*}=1.2 \mathrm{i}$ in the presence of the waveguide, as marked by the dashed line in Fig. \ref{fig4}B. This dipole can excite guided wave propagating unidirectionally in $+z$ direction in the silicon waveguide because $\kappa_{\mathrm{pw}}\left(+k_{\mathrm{wg}}\right) \gg \kappa_{\mathrm{pw}}\left(-k_{\mathrm{wg}}\right)$, where $\kappa_{\mathrm{pw}}\left(\pm k_{\mathrm{wg}}\right)$ is the coupling coefficient for the guided wave propagating in the $\pm z$ direction. The directionality is clearly observed in the $H_{y}$ field of the system for $\theta=15$ degrees and $\varphi=-5$ degrees, as shown in Fig. \ref{fig4}C. Figure \ref{fig4}D shows the distribution of $|H_{y}|$ inside the waveguide induced by the ideal circular dipole (solid blue line) and the optimized circular dipole (solid yellow line) of the helix, which has been normalized by the value of $|H_{y}|$ at $z=-1.5\lambda$ (corresponding to the amplitude of the guided wave propagating in $-z$ direction). We notice that the directionality, defined as $\left|H_{y}\left(+k_{\mathrm{wg}}\right) / H_{y}\left(-k_{\mathrm{wg}}\right)\right|$, reaches 52 for the optimized circular dipole and 10 for the ideal circular dipole. 

\begin{figure}[tp]
\centering
\includegraphics{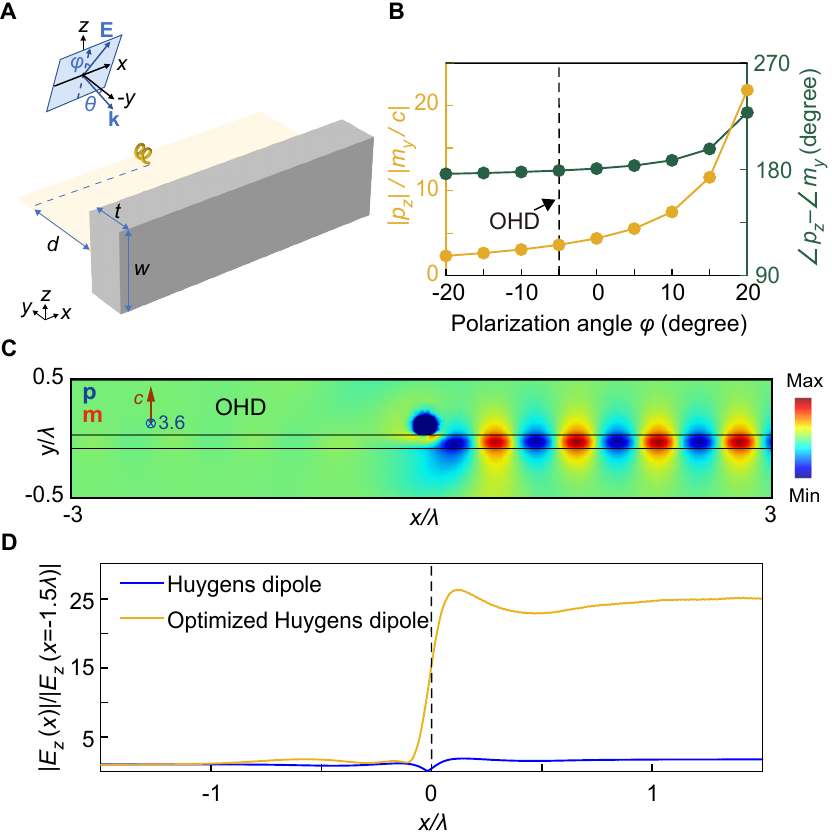}
\caption{\textbf{Directional excitation of guided wave by the Huygens-dipole face of the helix}. (\textbf{A}) Schematic of the helix-waveguide coupling configuration for demonstrating the directionality of the Huygens dipole.(\textbf{B}) The relative amplitude and phase of the dipoles $p_{z}$ and $m_{y}$ as a function of polarization angle $\varphi$. The incident angle is $\theta=5$ degrees. (\textbf{C}) Unidirectional coupling to the waveguide induced by the optimized Huygens dipole (OHD) in the helix. (\textbf{D}) The directionality of the ideal Huygens dipole and the optimized Huygens dipole in the helix.  } \label{fig5}
\end{figure}

To demonstrate the directional excitation of guided wave with the Huygens-dipole face of the DDD, we place the helix near the waveguide surface parallel to $zx$-plane with a distance of $80$ nm, as shown in Fig. \ref{fig5}A, which enables matching between the Huygens dipole $\mathbf{D}_{\text {Huy }}=\left(p_{z} \hat{\mathbf{e}}_{z}, m_{y} \hat{\mathbf{e}}_{y}\right)$ and the fields of the guided wave $\left(E_{z}, B_{y}\right)$. By tuning the incident angle $\theta$ and polarization angle $\varphi$ to satisfy $\left|p_{z} E_{z}^{*}+m_{y} B_{y}^{*}\right| \rightarrow 0$ for the guided wave propagating in $-x$ direction, we obtain the optimized Huygens dipole with $p_{z} /\left(m_{y} / c\right)=-c B_{y}^{*} / E_{z}^{*}=-3.6 \text { at } \theta=5 \text { degrees and } \varphi=-5 \text { degrees }$, as shown in Fig. \ref{fig5}B. Figure \ref{fig5}C shows the electric field $E_{z}$ of the guided wave excited by the optimized Huygens dipole, which propagates predominantly in $+x$ direction. In Fig. \ref{fig5}D, we plot the electric field inside the waveguide excited by the optimized Huygens dipole (solid yellow line), which has a directionality of 24 and is much larger than the directionality of the ideal Huygens dipole (solid blue line). 

\begin{figure}[t]
\centering
\includegraphics{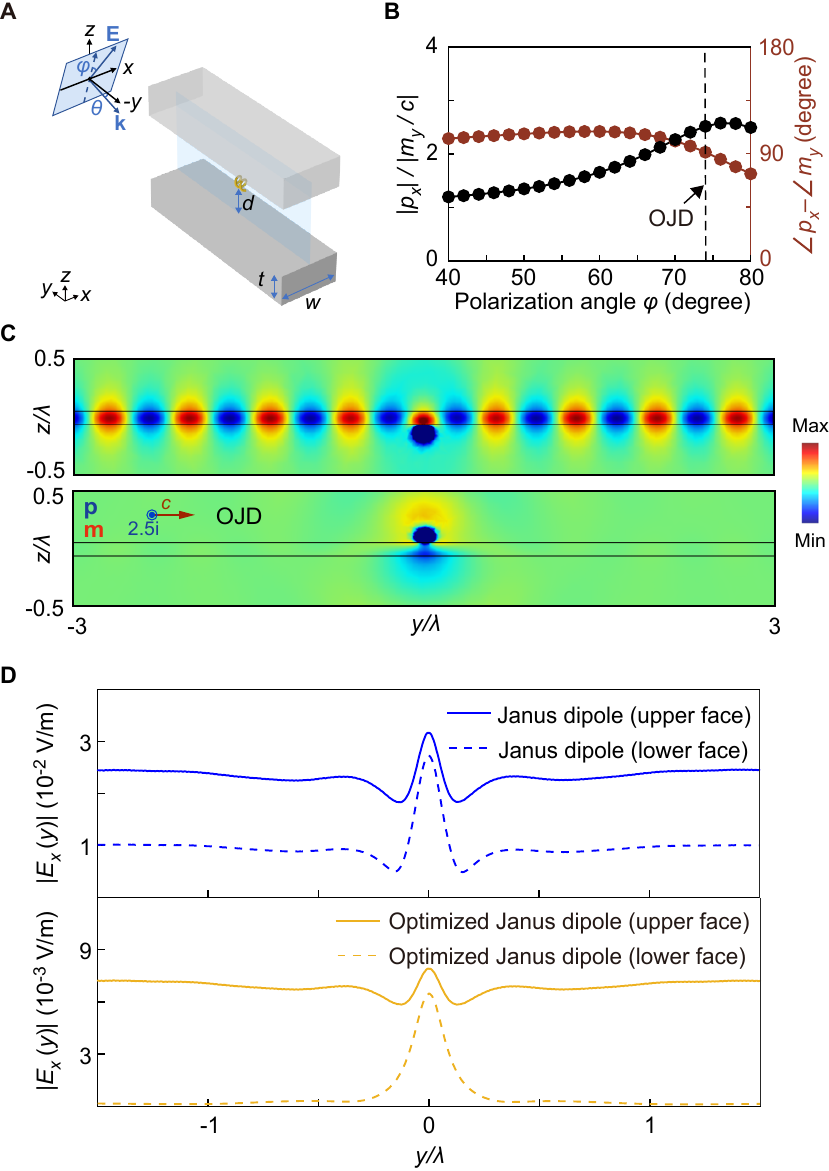}
\caption{\textbf{Directional excitation of guided wave by the Janus-dipole face of the helix}. (\textbf{A}) Schematic of the helix-waveguide coupling configuration for demonstrating the directionality of the Janus dipole. (\textbf{B}) The relative amplitude and phase of the dipoles $p_{x}$ and $m_{y}$ as a function of polarization angle $\varphi$. The incident angle is $\theta=75$ degrees. (\textbf{C}) When the optimized Janus dipole (OJD) of the helix is located below the waveguide, it can couple to the waveguide and excite the guided modes propagating in both directions. However, when the OJD is located above the waveguide, there is no coupling. (\textbf{D}) The directionality of the ideal Janus dipole (above) and the optimized Janus dipole (below) in the helix. } \label{fig6}
\end{figure}

To demonstrate the directional excitation of guided wave with the Janus-dipole face of the DDD, we place the helix $d=80$ nm above/below the waveguide surfaces parallel to $xy$-plane to turn on the directionality of $\mathbf{D}_{\text {Jan }}=\left(p_{x} \hat{\mathbf{e}}_{x}, m_{y} \hat{\mathbf{e}}_{y}\right)$, as shown in Fig. \ref{fig6}A. The directionality of the Janus dipole manifests as side-dependent coupling/noncoupling to the waveguide \cite{picardi2018janus}, i.e., whether it couples to the waveguide depends on which side of the helix is facing the waveguide. To achieve a high directionality, we optimize the Janus dipole by requiring $\left|p_{x} E_{x}^{*}+m_{y} B_{y}^{*}\right| \rightarrow 0$ for the coupling between the helix and waveguide, which gives $p_{x} /\left(m_{y} / c\right)=-c B_{y}^{*} / E_{x}^{*}=2.5 \mathrm{i}$ at incident angle $\theta=75$ degrees and polarization angle $\varphi=74$ degrees, as shown in Fig. \ref{fig6}B. Figures \ref{fig6}C and \ref{fig6}D show the electric field $E_{x}$ in the waveguide when the helix locates below and above the waveguide, respectively. As seen, the optimized Janus dipole of the helix only couples to the upper waveguide, corresponding to Fig. \ref{fig6}C. Figure \ref{fig6}D shows the electric field amplitude $|E_{x}|$ inside the waveguide for the two configurations in Fig. \ref{fig6}C. The solid (dashed) yellow line denotes the result for helix locating near the upper (lower) surface of the waveguide. The directionality in this case is defined as the ratio between the solid and dashed yellow lines, and it achieves a value of 40. For comparison, we also show the results of the ideal Janus dipole of the helix in Fig. \ref{fig6}D denoted by solid (dashed) blue lines, which has the directionality of 2.5. To further achieve unidirectional propagation of the excited guide wave in  $+y$ ($-y$) direction with the Janus dipole, we can simply add loss to one half of the the upper (lower) waveguide or truncate the waveguide (see Supplementary Fig. S6). 

The directionality of all three dipoles can be easily flipped by tuning the propagation and polarization directions of the incident plane wave (see the demonstration in Supplementary Figs. S3-S5). By combining the three configurations in Fig. 4A, Fig. 5A, and Fig. 6A, we obtain the high-dimensional and multifunctional system in Fig. \ref{fig1}. Demonstration of light routing with this system is shown in Supplementary Fig. S6. The amplitude of directionality can still reach around 10 in presence of the coupling among the waveguides, exhibiting robustness of the mechanism.

\begin{figure}[tp]
\centering
\includegraphics{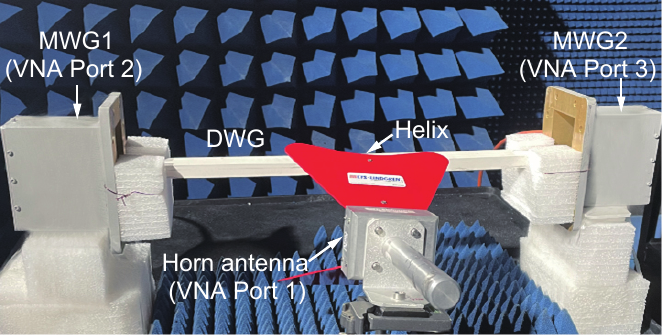}
\caption{\textbf{Experiment setup}. A horn antenna that excites the helix is mounted on a stand that allows azimuthal and elevation rotations. The dielectric waveguide (DWG) transitions into metallic waveguide (MWG) launchers on both ends. } \label{fig7}
\end{figure}

\begin{figure*}[ht]
\centering
\includegraphics{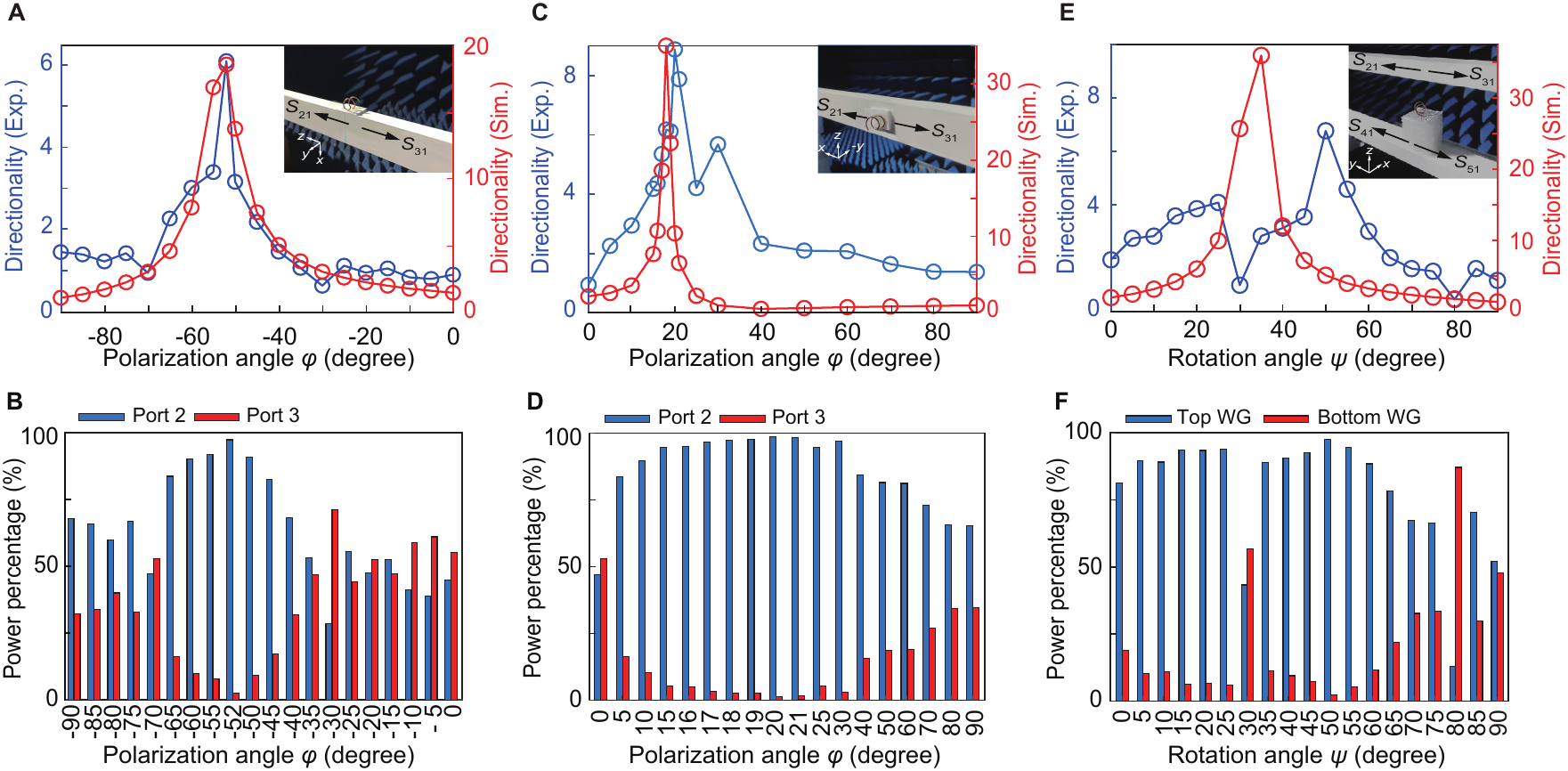}
\caption{\textbf{Experimental demonstrations of the directionality of the DDD in the helix}. (\textbf{A}, \textbf{B}) Simulated and experimental directionality of the circular-dipole face of the helix as a function of the polarization angle $\varphi$, as well as the power percentage of Port 2 and Port 3. (\textbf{C}, \textbf{D}) Simulated and experimental directionality of the Huygens-dipole face of the helix as a function of the polarization angle $\varphi$, as well as the power percentage of Port 2 and Port 3. (\textbf{E}, \textbf{F}) Simulated and experimental directionality of the Janus-dipole face of the helix as a function of the rotation angle $\psi$, as well as the power percentage of the top and bottom waveguides. } \label{fig8}
\end{figure*}  

\subsection*{Microwave experiments} 
We experimentally demonstrate the fascinating directional scattering and coupling of the DDD in the microwave regime. A photo of the experimental setup is shown in Figure \ref{fig7}. We use a copper helix supporting the dipole resonance at 2.35 GHz. A horn antenna connected to Port 1 of a Vector Network Analyzer (VNA) provides the plane wave excitation. The antenna is mounted on a stand that allows angular rotations amounting to variations in the propagation and polarization directions of the incident wave (i.e., $\theta$ and $\varphi$). The plane wave excites the appropriate dipoles of the helix, which couple waves into the dielectric waveguide in a directional manner. Both ends of the dielectric waveguide feature tapers to optimize coupling into open-ended WR-430 metallic waveguide launchers, which are connected to Ports 2 and 3 of the VNA. This setup enables us to measure the coupling strength by comparing the power sent through Port 1 to the powers received at Ports 2 and 3 respectively. By tuning the directions of incidence and polarization, the helix can operate as a DDD to realize near-field directional waveguiding.

In the circular-dipole experiment, the helix-waveguide configuration follows the settings in Fig. \ref{fig4}A. The helix is placed near the dielectric waveguide with the gap distance of 2 mm, as shown in the inset of Fig. \ref{fig8}A. The symbol lines in Fig. \ref{fig8}A show the simulated and experimental amplitude directionality as a function of the polarization angle $\varphi$ when $\theta=7$ degrees. As seen, the helix can numerically realize an optimized circular dipole with the directionality of 18.5 at $(\theta, \varphi)=(7$ degrees, $-52$ degrees). The experimental directionality, given by the ratio of the received amplitude signals at Ports 2 and 3 $(\left.\left|\mathbf{E}_{2}^{\text {out }}\right| /\left|\mathbf{E}_{3}^{\text {out }}\right|=\left|S_{21}\right| /\left|S_{31}\right|\right)$ exhibits a peak at $\varphi=-52$ degrees. The high directionality value of 6.1 indicates the achievement of superior coupling to Port 2, which accounts for $97.4 \%$ of the total coupled power as shown in Fig. \ref{fig8}B. For the Huygens-dipole experiment, we follow the configuration settings in Fig. \ref{fig5}A and place the helix near the dielectric waveguide with the gap distance of 2 mm as shown in the inset of Fig. \ref{fig8}C. We again plot the simulated and measured directionality as a function of $\varphi$ (for $\theta=7$ degrees) in Fig. \ref{fig8}C. As seen, the helix can numerically realize an optimized Huygens dipole at $(\theta, \varphi)=(7$ degrees, $18$ degrees), with a simulated directionality of 35. The experimental directionality $(\left.\left|\mathbf{E}_{2}^{\text {out }}\right| /\left|\mathbf{E}_{3}^{\text {out }}\right|=\left|S_{21}\right| /\left|S_{31}\right|\right)$ reaches a peak value of 8.9 at $\varphi=20$ degrees, where $98.75 \%$ of the coupled power is received at Port 2 as shown in Fig. \ref{fig8}D. For the circular and Huygens dipoles of the DDD, the simulation and experiment results show great agreement, which verifies the robustness of our proposed mechanism.

In a slight deviation from the settings in Fig. \ref{fig6}A, in order to overcome a wave blockage phenomenon, we adopt an alternative excitation geometry in demonstrating the side-dependent coupling property of the Janus dipole (see Supplementary Fig. S8). Figure \ref{fig8}E shows the placement of the helix midway between the two dielectric waveguides with gap distance 50 mm. In this approach, the incident direction of the plane wave (the wavevector $\mathbf{k}$) is on the $xy$ plane, such that the line of sight from the feed antenna to the helix is unobstructed by the waveguides. The directionality optimization of Janus dipole can be achieved by tuning the incident angle $\theta$ (the angle between $\mathbf{k}$ and $-y$) and rotation angle $\psi$ (the angle between the helix axis and $y$ direction) on the $xy$ plane. We divide the experiment into two parts: (1) the coupling between the helix and upper waveguide, and (2) the coupling between the helix and lower waveguide. Figure \ref{fig8}E shows the simulated and measured amplitude directionality of the Janus dipole as a function of $\psi$ when $\theta=40$ degrees, where the directionality is defined as the ratio of the field coupled to the top waveguide (Port 2 + Port 3) and bottom waveguide (Port 4 + Port 5). As seen, the helix can numerically realize an optimized Janus dipole with the directionality of 36 at $(\theta, \psi)=(40$ degrees, $35$ degrees). The experimental directionality, given by $\left(\left|\mathbf{E}_{2}^{\text {out }}\right|+\left|\mathbf{E}_{3}^{\text {out }}\right|\right) /\left(\left|\mathbf{E}_{4}^{\text {out }}\right|+\left|\mathbf{E}_{5}^{\text {out }}\right|\right)=\left(\left|S_{21}\right|+\left|S_{31}\right|\right) /\left(\left|S_{41}\right|+\left|S_{51}\right|\right)$, reaches a peak with the value of 6.8 at $\psi=50$ degrees. At this directionality, nearly $98 \%$ of the power couples to the top waveguide as shown in Fig. \ref{fig8}F. The simulation and experiment results show consistent trend but have a 15-degree shift in the optimal orientation angle, likely due to slight inaccuracies in the measurement setup. 

\section*{\textcolor[RGB]{200,36,38}{DISCUSSION}}
\noindent In conclusion, we theoretically and experimentally demonstrate that the synergy of chirality and anisotropy can enable the realization of the DDD to give the circular dipole, Huygens dipole, and Janus dipole under the excitation of a linearly polarized plane wave at the same frequency. We apply the three directional dipoles on different faces of the DDD to realize complete control of directional optical coupling to the dielectric waveguide in the infrared regime. The phenomena are experimentally verified in the microwave regime using a copper helix placed near dielectric waveguides. The emergence of the directional dipoles in the helix particle can be well understood based on a mode expansion theory. It is found that chirality gives rise to the Janus dipole via the odd-order eigenmodes of the helix; anisotropy gives rise to the circular dipole via both the odd-order and even-order eigenmodes; anisotropy and chirality together generate the Huygens dipole via the odd-order and even-order eigenmodes. The analytical theory enables straightforward designing of optimal multifunctioned directional sources for arbitrary waveguides. The proposed DDD can enable the multiplexed control of near-field and far-field directionality via three different degrees of freedom: spin, power flow, and reactive power. The circular-dipole face and the Janus-dipole face of the DDD can be employed to achieve light routing in photonic integrated circuits and to realize asymmetric coupling of resonators for exploring non-Hermitian physics. The Huygens-dipole face of the DDD can find essential applications in designing highly directional quantum sources and nonreflecting metasurfaces. The unified realization and control of all three types of directionalities in orthogonal directions opens numerous opportunities for realizing multiple functionalities in the high-dimensional space by a single meta-atom or metasurface, where the circular, Huygens, and Janus directional properties can be freely switched upon different incidence. A meta-structure of this element can also be constructed with multiple resonances intertwined to couple together different types of directionality, which can further enrich the ability of light manipulation. The possibilities are vast. The mechanism may also be extended to other classical waves such as sound waves.

\section*{\textcolor[RGB]{200,36,38}{MATERIALS AND METHODS}}
\subsection*{Mode expansion theory}
The eigenmodes of the metal helix can be determined by mapping the eigenmodes of the corresponding nanorod, which can be semi-analytically determined as follows. The eigenmodes of the nanorod are Fabry-Perot standing waves of currents with eigenfrequencies $\omega_n$ satisfying \cite{novotny2012principles}:
\begin{equation}
    \qquad \omega_{n} \sqrt{\mu_{0} \varepsilon_{0}} n_{\text {eff}}(\omega) L+\Phi(\omega)=n \pi,
     \label{eq:11}
\end{equation}
where $L=N \sqrt{4 \pi^{2} R^{2}+P^{2}}$ is the arc length of the $N-$turn helix, $n$ is the order of the eigenmodes, $n_{\text {eff }}(\omega)$ is the effective refractive index of propagating surface plasmon, and $\Phi(\omega)$ is the reflection phase at the ends of the helix. The values of $n_{\text {eff }}(\omega)$ and $\Phi(\omega)$ are approximately constant for a fixed $r$ over a range of frequencies, and they can be numerically determined via solving for the $\mathrm{TM}_{0}$ guided mode of the nanorod \cite{novotny2007effective}. Equation (\ref{eq:11}) allows us to analytically calculate the eigenfrequency $\omega_{n}$
and the propagating constant $\gamma_{n}(\omega)=\omega_{n} \sqrt{\mu_{0} \varepsilon_{0}} n_{\text {eff }}(\omega)$ of the current wave. The currents on the nanorod can be approximately expressed as $\cos \left(\gamma_{n} l\right)$ for odd orders and $\sin \left(\gamma_{n} l\right)$ for even orders with $l \in[-L / 2, L / 2]$. The eigen currents of the helix can be obtained by mapping these currents of the nanorod as 
\begin{equation}
     \mathbf{J}_{n}=\left\{\begin{array}{l}
\cos \left(\gamma_{n} l\right) \mathbf{t}, n=1,3,5, \ldots \\
\sin \left(\gamma_{n} l\right) \mathbf{t}, n=2,4,6, \ldots
\end{array}\right. 
\label{eq:12}
\end{equation}
with $\mathbf{t}=(\frac{\sigma 2 \pi N R}{L} \cos (\frac{2 \pi N l}{L}+N \pi), \frac{N P}{L}, \frac{2 \pi N R}{L} \sin (\frac{2 \pi N l}{L}+N \pi))$ being the tangent direction vector of the helix path. The Green’s function of the helix can be constructed as $\mathbf{G}\left(\mathbf{r}, \mathbf{r}^{\prime}\right)=\sum_{n} \frac{\mathbf{J}_{n}(\mathbf{r}) \mathbf{J}_{n}^{*}\left(\mathbf{r}^{\prime}\right)}{A_{n}\left(\omega_{n}^{2}-\omega^{2}-i \omega \gamma\right) \omega^{2}}=\sum_{n} a_{n}(\omega) \mathbf{J}_{n}(\mathbf{r}) \mathbf{J}_{n}^{*}\left(\mathbf{r}^{\prime}\right)$ \cite{novotny2012principles}, where $A_{n}$ is the normalization coefficient of the eigen current $\mathbf{J}_{n}$ at eigenfrequency $\omega_{n}$. The expansion coefficient $a_{n}(\omega)$ contains holistic resonance characteristics of the helix and is independent of the excitation. It can be numerically determined based on the induced current density 
\begin{equation}
\begin{gathered}
\mathbf{J}(\mathbf{r})=-\mathrm{i} \omega^{3} \varepsilon_{0}^{2}\left(\Delta \varepsilon_{r}\right)^{2} \mu \int \mathbf{G}\left(\mathbf{r}, \mathbf{r}^{\prime}\right) \cdot \mathbf{E}\left(\mathbf{r}^{\prime}\right) \mathrm{d} V_{\mathrm{p}} \\
=\sum_{n} B_{n} a_{n} \mathbf{J}_{n}(\mathbf{r}),
\label{eq:13}
\end{gathered}
\end{equation}
where
$$
  B_{n}=-\mathrm{i} \omega^{3} \varepsilon_{0}^{2}\left(\Delta \varepsilon_{r}\right)^{2} \mu \int \mathbf{J}_{n}^{*}\left(\mathbf{r}^{\prime}\right) \cdot \mathbf{E}\left(\mathbf{r}^{\prime}\right) \mathrm{d} V_{\mathrm{p}}.
$$
Here, $\Delta \varepsilon_{r}=\varepsilon_{\mathrm{Au}}-1$ is the relative permittivity contrast between the helix particle and the background medium (i.e., free space). The integral $\int \mathbf{J}_{n}^{*}\left(\mathbf{r}^{\prime}\right) \cdot \mathbf{E}\left(\mathbf{r}^{\prime}\right) \mathrm{d} V_{\mathrm{p}}$ is evaluated over the volume of the helix particle 
$V_{\mathrm{p}}$. The coefficient $a_{n}$ can be determined after one simulation of the induced current, with which we then can analytically calculate the induced dipoles for any excitations as 
\begin{equation}
    \mathbf{p}=\frac{\mathrm{i}}{\omega} \int \mathbf{J}(\mathbf{r}) \mathrm{d} V_{\mathrm{p}}=\sum_{n} B_{n} a_{n} \mathbf{p}_{n}
    \label{eq:14}
\end{equation}
and
\begin{equation}
   \mathbf{m}=\frac{1}{2} \int \mathbf{r} \times \mathbf{J}(\mathbf{r}) \mathrm{d} V_{\mathrm{p}}=\sum_{n} B_{n} a_{n} \mathbf{m}_{n},
    \label{eq:15}
\end{equation}
where $\mathbf{p}_{n}=\frac{\mathrm{i}}{\omega} \int \mathbf{J}_{n}(\mathbf{r}) \mathrm{d} V_{\mathrm{p}}$ and $\mathbf{m}_{n}=\frac{1}{2} \int \mathbf{r} \times \mathbf{J}_{n}(\mathbf{r}) \mathrm{d} V_{\mathrm{p}}$ are dipoles attributed to the eigen current $\mathbf{J}_{n}$. For odd values of $n$ in Eq. (\ref{eq:12}), the dominant dipole components are
\begin{equation}
    \begin{gathered}
\left(\mathbf{p}_{n}\right)_{x}=\frac{\mathrm{i} \sigma C R}{\omega}\left(\frac{\sin {\phi_{n}}}{\gamma_{n}-K}+\frac{\sin {\phi_{n}}}{\gamma_{n}+K}\right),\\
\left(\mathbf{p}_{n}\right)_{y}=\frac{\mathrm{i} C P \sin {\phi_{n}}}{\pi \omega \gamma_{n}},\left(\mathbf{p}_{n}\right)_{z}=0,\\
\left(\mathbf{m}_{n}\right)_{y}=\frac{-\sigma C R^{2} \sin {\phi_{n}}}{\gamma_{n}} ,
\label{eq:16}
\end{gathered}
\end{equation}
where $C=2 N \pi^{2} r^{2} / L$, $\phi_{n}=\gamma_{n} L / 2$, $K=2 \pi N / L$, $N$ is the number of turns of the helix, $L$ is arc length of the helix, and $\sigma=+ 1$ ($\sigma=-1$) for the left (right) handed helix.  For even values of $n$ in Eq. (\ref{eq:12}), the dominant dipole components are  
\begin{equation}
\begin{gathered}
\left(\mathbf{p}_{n}\right)_{x}=0,\left(\mathbf{p}_{n}\right)_{y}=0,\\
\left(\mathbf{p}_{n}\right)_{z}=\frac{\mathrm{i} C R}{\omega}\left(\frac{\sin \phi_{n}}{\gamma_{n}-K}-\frac{\sin \phi_{n}}{\gamma_{n}+K}\right),\\
\left(\mathbf{m}_{n}\right)_{y}=0 .
\label{eq:17}
\end{gathered}
\end{equation}

\subsection*{Numerical simulations}
All the full-wave numerical simulations are performed with the package COMSOL Multiphysics (www.comsol.com). For the simulation of the gold helix at optical frequencies, we set the pitch $P=75$ nm, outer radius $R=46$ nm, and inner radius $r=11$ nm (refer to the inset in Fig. \ref{fig2}A for definition of the geometric parameters). The center axis of the helix is along $y$ direction. The relative permittivity of the gold helix is characterized by the Drude model $\varepsilon_{\mathrm{Au}}=1-\omega_{\mathrm{p}}^{2} /\left(\omega^{2}+\mathrm{i} \omega \omega_{\mathrm{t}}\right)$, where $\omega_{\mathrm{p}}=1.36 \times 10^{16}     \mathrm{rad} / \mathrm{s} \text { and } \omega_{\mathrm{t}}=7.1 \times 10^{13}    \mathrm{rad} / \mathrm{s}$ \cite{olmon2012optical}. The silicon waveguide (relative permittivity $\varepsilon_{\mathrm{Si}}=12$) in Figs. \ref{fig4}, \ref{fig5}, and \ref{fig6} has a rectangular cross section of $w \times t=620 \mathrm{~nm} \times 310 \mathrm{~nm}$. In the simulations of the directional excitations of guided wave, we apply absorption boundary conditions on both ends of the waveguide to suppress any reflections. Beyond the waveguide region, an open boundary condition is applied.  

\subsection*{Experiments}
The helix is made of copper with the electrical conductivity $\sigma=5.813 \times 10^{7} \mathrm{~S} / \mathrm{m}$. It has pitch $P=8$ mm, outer radius $R=5.5$ mm, and inner radius $r=0.5$ mm. The dielectric waveguide with $\varepsilon_{r}=12$ was fabricated by CNC technology. It has a cross-sectional dimension of $30 \mathrm{~mm} \times 15 \mathrm{~mm}$ and a length of 620 mm. The length of the taper at both ends of the waveguide is 60 mm. The experimental setup shown in Fig. \ref{fig7} can be considered a 3-port network. The couplings from the source antenna (Port 1) to the waveguide outputs (Ports 2 and 3) via the helix can be directly determined by measuring the S-parameters $S_{21}$ and $S_{31}$. The Janus system in Fig. \ref{fig8}E corresponds to a 5-port network, where the couplings to waveguide outputs (Ports 2-5) were determined similarly by measuring the S-parameters $S_{21},S_{31},S_{41}$ and $S_{51}$. A background measurement is taken whereby $[S]_{ \text { bkgd }}$ is measured when the helix is absent from the experimental setup. This measurement picks up faint spurious signals directly coupled from the source antenna to the waveguide outputs or scattered by other objects within the measurement chamber. Performing a background calibration $[S]=[S]_{\text {helix}}-[S]_{\text {no helix}}$ minimizes these background contributions to the measurement.

\def\bibsection{}
\section*{\textcolor[RGB]{200,36,38}{REFERENCES AND NOTES}}
\renewcommand\bibnumfmt[1]{#1.}
\nocite{trinh1980metal,dudorov2002rectangular}
\bibliography{MyCollection}

\noindent
\\
\textbf{\sffamily{Acknowledgements:}} We thank J. Peng for useful discussions. \textbf{\sffamily{Funding:}} The work described in this paper was supported by the National Natural Science Foundation of China (No. 11904306) and grants from the Research Grants Council of the Hong Kong Special Administrative Region, China (Projects No. C6013-18G and No. CityU 11301820). \textbf{\sffamily{Author contributions:}} S.W. conceived the project. Y.C. conducted the numerical simulations and developed the theory. K.A.O. and B.X. performed the microwave experiments. D.L. assisted in theoretical analysis. Y.C., K.A.O., and S.W. wrote the draft. S.W. and A.M.H.W. supervised the project. All authors contributed to discussions and the polishing of the manuscript. \textbf{\sffamily{Competing interests:}} The authors declare no competing interests. \textbf{\sffamily{Data and materials availability:}} All data needed to evaluate the conclusions in the paper are present in the paper and/or the Supplementary Materials.

\clearpage
\section*{\textcolor[RGB]{200,36,38}{SUPPLEMENTARY MATERIALS}}
\vspace{0.05in}
\subsection*{Eigenmodes of the nanorod}
\setcounter{figure}{0}
\renewcommand{\figurename}{\textbf{Fig.}}
\renewcommand{\thefigure}{\textbf{S\arabic{figure}}}

Figure \ref{figS1} shows the eigenfrequency $\omega_{n}$ of the gold nanorod ($r=11$ nm) for the first eigenmode $n=1$ and propagation constant $\gamma_{n}$ of the nanorod current for the eigenmodes of orders $n=1,2,3,4$ under different lengths. The cross symbols denote the analytical results obtained using Eq. (11) (see Materials and Methods), which agree well with the numerical results obtained using COMSOL (denoted by the circle symbols).
\begin{figure}[htp]
\centering
\includegraphics[scale=1]{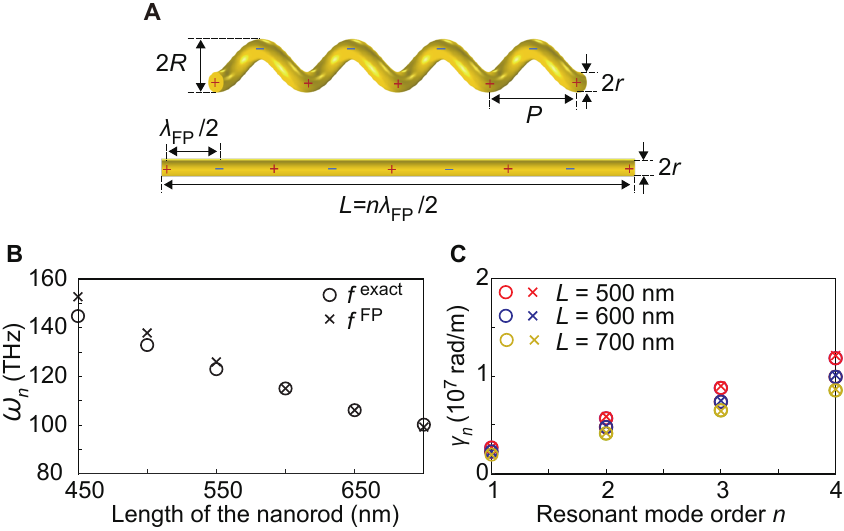}
\caption{\textbf{Eigenmode properties of the nanorod}. (\textbf{A}) Charge distribution for the Fabry-Perot standing wave of order $n$ in the helix and the nanorod. (\textbf{B}) Comparison between analytical results (cross) and numerical results (circle) of the eigen frequency for different lengths of the nanorod. (\textbf{C}) Comparison of the propagation constants of the eigenmodes obtained by the analytical method (cross) and numerical method (circle). }
\label{figS1}
\end{figure}

\subsection*{Alternative design of the helix as the DDD}
The physical mechanism of realizing the directional dipoles in an anisotropic chiral particle is robust, and the phenomena can be demonstrated in different designs of the helix working at different frequencies. In addition to the design in the main text, another design of the helix is shown in Fig. \ref{figS2}A, where the pitch is $P=90$ nm, the outer radius is $R=40$ nm, and the inner radius is $r=15$ nm. This helix works at the dipole resonant frequency of 147 THz, and it can give rise to the three types of directional dipole when the incident angle is 15 degrees, as shown in Fig. \ref{figS2}B. 

\begin{figure}[htp]
\centering
\includegraphics[scale=0.95]{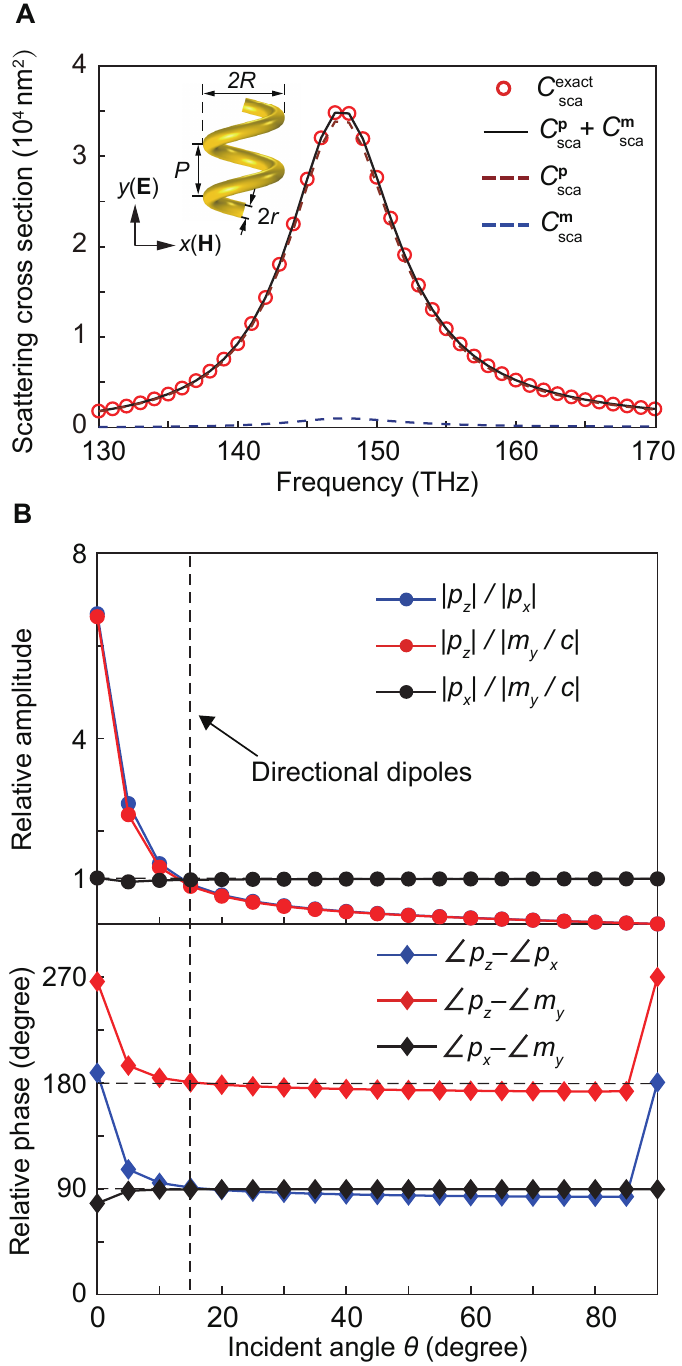}
\caption{\textbf{Electric and magnetic dipole moments induced in the helix}. (\textbf{A}) Scattering cross sections of the helix particle and the contributions of the electric and magnetic dipoles. (\textbf{B}) The relative amplitudes and phases of the dipole components as a function of the incident angle of the linearly polarized plane wave. The dashed line marks the parameters giving three directional dipoles.} \label{figS2}
\end{figure}

\subsection*{Directionality flippling of the DDD}

The directionality of the DDD can be flipped by activating the remaining three directional-dipole faces pointing in $-x$, $+y$, and $-z$ directions respectively. Following the settings of the helix-waveguide configurations in Figs. 4-6 (see Results), we reverse the directionality of all types of directional dipoles by tuning the incidence. As seen in Figs. \ref{figS3}A and \ref{figS3}B, when the incident angle $\theta=165$ degrees and polarization angle $\varphi=-5$ degrees, the helix can serve as an optimized circular dipole ($p_{z} / p_{x}=-1.2 \mathrm{i}$) with the directionality of 32 pointing in $-z$ direction, which can be clearly observed in the $H_{y}$ field of the system in Fig. \ref{figS3}C. For the Huygens-dipole face, when $\theta=175$ degree and $\varphi=-5$ degrees, the helix can realize an optimized Huygens dipole ($p_{z} / (m_{y}/c)=3.6$) with the directionality of 38 pointing in $-x$ direction as shown in Figs. \ref{figS4}A and \ref{figS4}B. Fig. \ref{figS4}C clearly exhibits the unidirectional excitation of the $-x$-propagating guided wave. Figure \ref{figS5} shows the reversed directionality of Janus-dipole face. When $\theta=-28$ degrees and $\varphi=-68$ degrees, the helix can generate the optimized Janus dipole ($p_{x} / (m_{y}/c)=-2.5 \mathrm{i}$) and predominantly couples light to the lower waveguide with a directionality of 22. It can also be clearly observed in the electric field distribution ($E_{x}$) of the system in Fig. \ref{figS5}C. 
\begin{figure}[htp]
\centering
\includegraphics[scale=1]{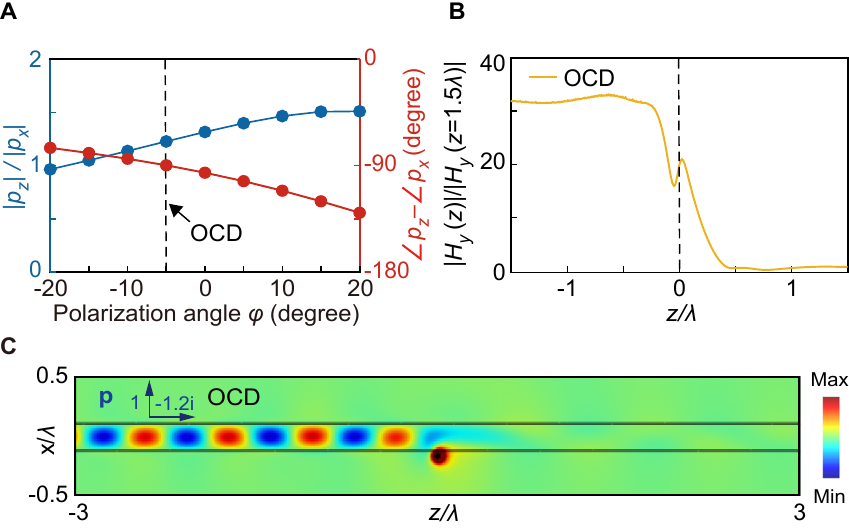}
\caption{\textbf{Reversed directionality of circular-dipole face}. (\textbf{A}) The relative amplitude and phase of the dipoles $p_{z}$ and $p_{x}$ as a function of polarization angle $\varphi$. The incident angle is $\theta=165$ degrees. (\textbf{B}) The directionality of optimized circular dipole (OCD) and (\textbf{C}) the unidirectional excitation of guided waves propagating in $-z$ direction. } \label{figS3}
\end{figure}

\begin{figure}[htp]
\centering
\includegraphics[scale=1]{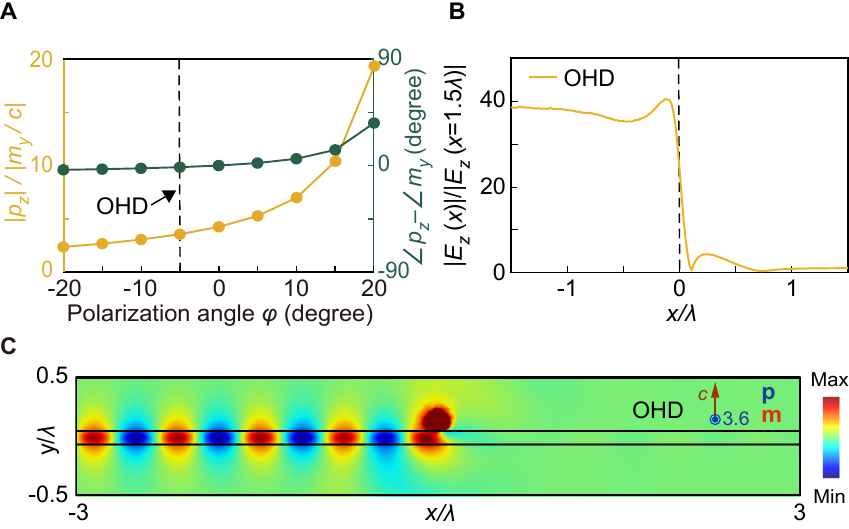}
\caption{\textbf{Reversed directionality of Huygens-dipole face}. (\textbf{A}) The relative amplitude and phase of the dipoles $p_{z}$ and $m_{y}$ as a function of polarization angle $\varphi$. The incident angle is $\theta=175$ degrees. (\textbf{B}) The directionality of optimized Huygens dipole (OHD) and (\textbf{C}) the unidirectional excitation of guided waves propagating in $-x$ direction. } \label{figS4}
\end{figure}

\begin{figure}[htp]
\centering
\includegraphics[scale=1]{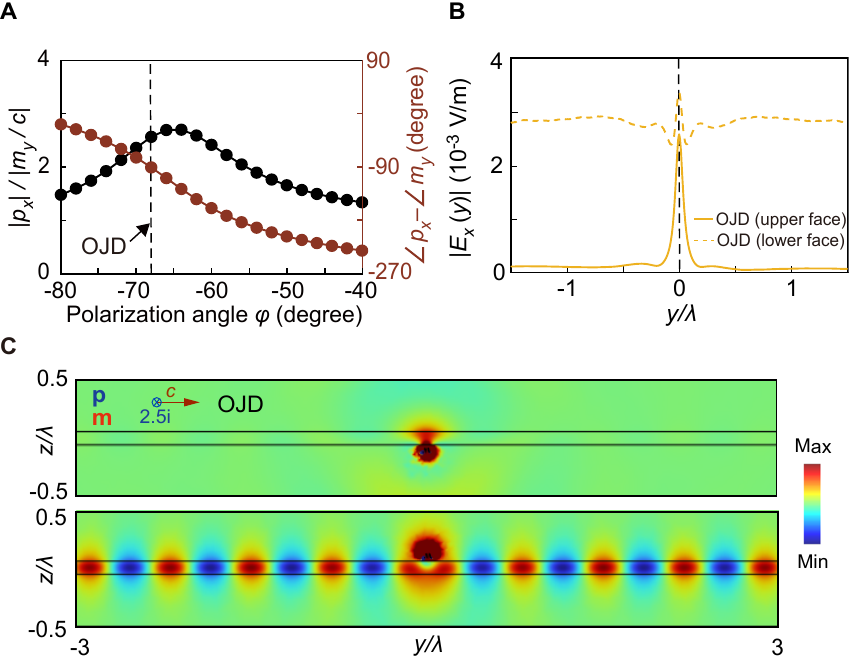}
\caption{\textbf{Reversed directionality of Janus-dipole face}. (\textbf{A}) The relative amplitude and phase of the dipoles $p_{x}$ and $m_{y}$ as a function of polarization angle $\varphi$. The incident angle is $\theta=-28$ degrees. (\textbf{B}) The directionality of optimized Janus dipole (OJD) and (\textbf{C}) the unidirectional coupling to the lower waveguide. } \label{figS5}
\end{figure}
\subsection*{High-dimensional directional system}

To unify all types of directionality in one space and realize the switching among them, we construct a high-dimensional directional system as shown in Fig. \ref{figS6}A. Following the schematic in Fig. 1, the helix is surrounded by three sets of waveguide channels with the gap distance $d=925$ nm, where the transparent parts of the waveguide contain loss. The amplitude ratio of the guided waves in $\pm z$-, $\pm x$-, and $\pm y$-propagating channels corresponds to the directionality of circular dipole, Huygens dipole, and Janus dipole, respectively. The optimization of each dipole and the directional coupling will be influenced by the couplings between the nearby waveguides. By controlling the propagation and polarization directions (i.e., $\theta$ and $\varphi$) of the incident plane wave, we can achieve complete manipulation of directionality via different directional-dipole faces in orthogonal orientations. In Fig. \ref{figS6}B, when $\theta=8$ degrees and $\varphi=-5$ degrees, we achieve the circular dipole-type directionality with $\mathbf{D}_{\mathrm{cir}}^{\mathrm{e}}=\left(p_{x} \hat{\mathbf{e}}_{x}, p_{z} \hat{\mathbf{e}}_{z}\right)$ in the helix and guide light predominantly propagating in $+z$ direction. The optimized directionality by the dipole $p_{z}/p_{x} \approx -0.6+2.1\mathrm{i}$ can be clearly observed in the electric field distribution of the system ($zy$-view) in the inset, which reaches the value of almost 13. Similarly, when $\theta=20$ degrees and $\varphi=-25$ degrees, we realize the Huygens dipole-type directionality with $\mathbf{D}_{\mathrm{Huy}}=\left(p_{z} \hat{\mathbf{e}}_{z}, m_{y} \hat{\mathbf{e}}_{y}\right)$ in the helix and guide light predominantly propagating in $+x$ direction, as shown in Fig. \ref{figS6}C. From the inset, the optimized directionality by the dipole $p_{z} / (m_{y}/c) \approx -0.75$, described by the electric field distribution on the $xy$- plane, reaches about 8. As shown in Fig. \ref{figS6}D, when $\theta=-36$ degrees and $\varphi=-67$ degrees, we can achieve the Janus dipole-type directionality with $\mathbf{D}_{\mathrm{Jan}}=\left(p_{x} \hat{\mathbf{e}}_{x}, m_{y} \hat{\mathbf{e}}_{y}\right)$ in the helix and guide light predominantly propagating along $-y$ direction in the lower waveguide. The inset clearly shows the optimized directionality by the dipole $p_{x} / (m_{y}/c) \approx -1-1.4\mathrm{i}$ via the electric field distribution of the system ($yz$-view), which reaches the value of nearly 9. An alternative way for the Janus-dipole face to demonstrate the directionality is to truncate the waveguide at an appropriate position of the transparent part. When the reflected wave from the termination constructively interferes with the wave traveling in $+y$ ($-y$) direction in the upper (lower) waveguide, one can also achieve unidirectional propagation of the guided wave in $+y$ or $-y$ direction. Meanwhile, this method can improve the transmission efficiency by harvesting light into one waveguide channel.
\clearpage
\onecolumngrid

\begin{figure}[h]
\centering
\includegraphics[scale=0.95]{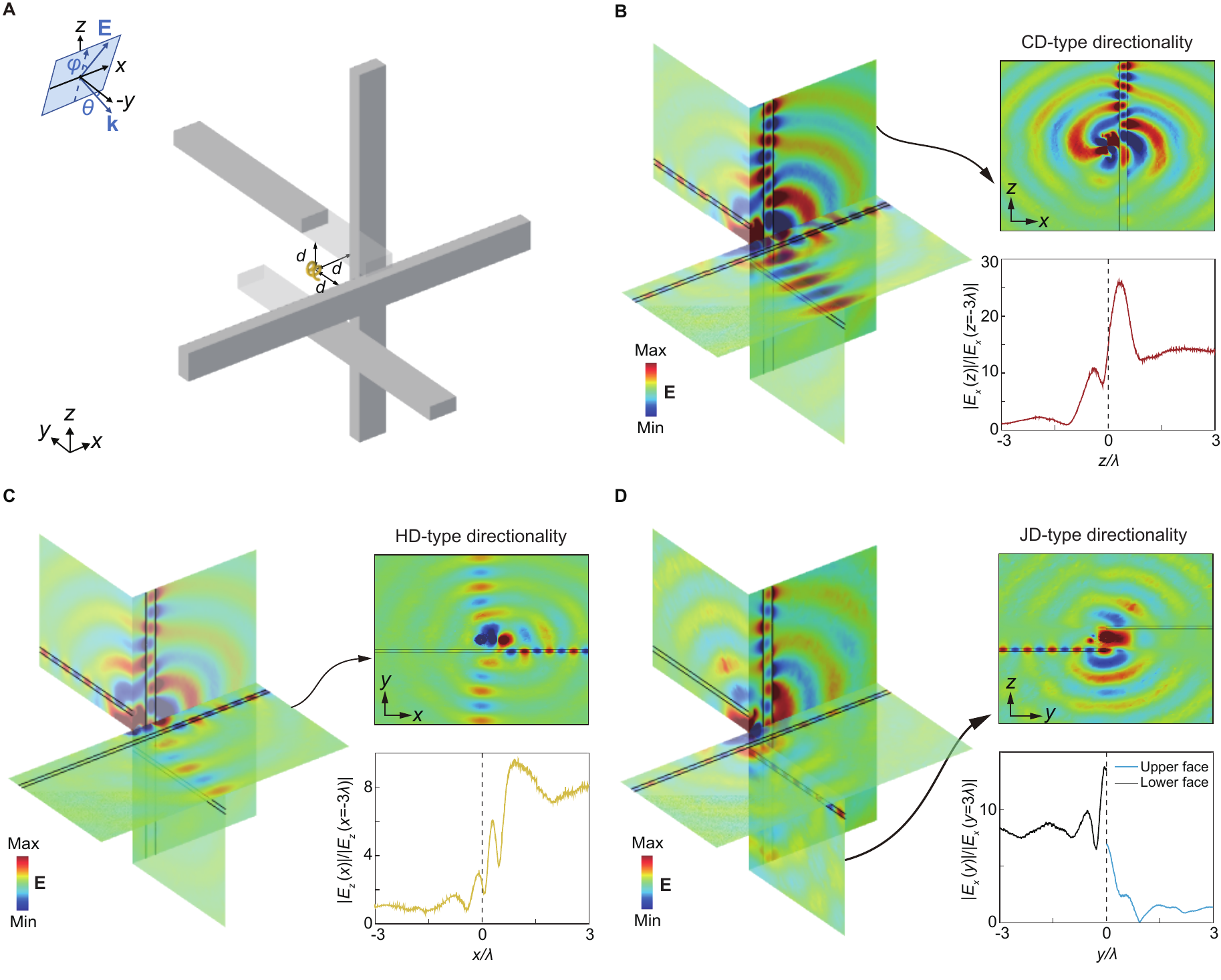}
\caption{\textbf{High-dimensional and multifunctional control of directionality}. (\textbf{A}) Schematic of high-dimensional directional system by the DDD. (\textbf{B}) Electric field distribution of the system at $\theta=8$ degrees and $\varphi=-5$ degrees for circular dipole (CD)-type directionality in $+z$ direction. The inset shows the optimized unidirectional coupling on the $zx$- plane as well as the directionality. (\textbf{C}) Electric field distribution of the system at $\theta=20$ degrees and $\varphi=-25$ degrees for Huygens dipole (HD)-type directionality in $+x$ direction. The inset shows the optimized unidirectional coupling on the $xy$- plane as well as the directionality. (\textbf{D}) Electric field distribution of the system at $\theta=-36$ degrees and $\varphi=-67$ degrees for Janus dipole (JD)-type directionality in $-y$ direction. The inset shows the optimized unidirectional coupling on the $yz$- plane as well as the directionality.} \label{figS6}
\end{figure}

\clearpage
\twocolumngrid

\subsection*{Dielectric waveguide to metallic waveguide transition}

The dielectric waveguide $\left(\mu_{r}=1, \varepsilon_{r}=12\right)$ used in this experiment with cross-sectional dimension of $30 \mathrm{~mm} \times 15 \mathrm{~mm}$ and a total length of 620 mm was fabricated by CNC technology. In opposition to what is obtainable in a metallic waveguide, the fields are not totally confined to the guiding structure in a dielectric waveguide. Transitioning between these two types of waveguides thus presents an interesting challenge. Many transition methods have been proposed over the years \cite{trinh1980metal,dudorov2002rectangular}. In this experiment, we adopt the end tapering method to ensure optimum transition from the dielectric waveguide to the standard WR430 metallic waveguide launchers, which are connected to the vector network analyzer. The two ends of the dielectric waveguide thus have a 60 mm-long taper.

\begin{figure}[htp]
\centering
\includegraphics[scale=0.75]{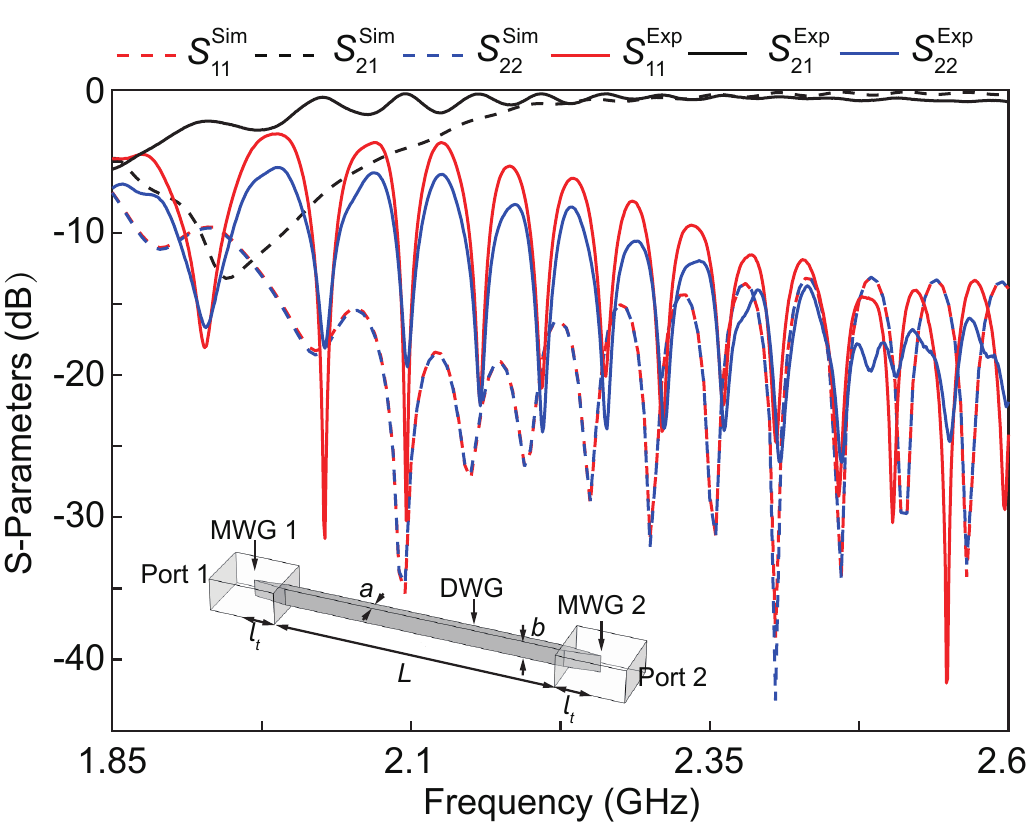}
\caption{\textbf{Dielectric waveguide to metallic waveguide transition}. The inset shows the simulation and experiment model (DWG – Dielectric waveguide with dimensions $a=15$ mm, $b=30$ mm, $l_{t}=60$ mm, $L=500$ mm, MWG – Metallic waveguide). Extracted simulation and experiment results show minimal reflection and excellent transmission from the input port (Port 1) to the output port (Port 2). } \label{figS7}
\end{figure}

An HFSS simulation setup to confirm the efficient transition is shown in the inset of Fig. \ref{figS7}. We excite Port 1 and measure the received power at Port 2 ($S_{21}$) as well as the reflected power back into Port 1, comprising the reflection from the metallic waveguide launcher (at Port 1) and the reflection from the end of the dielectric waveguide (at Port 2) back into Port 1 ($S_{11}$). We also experimentally extract the reflection and transmission parameters when the two ends of the dielectric waveguide are terminated with WR430 metallic waveguide launchers. The WR430 waveguides launchers have a coaxial-to-waveguide adaptor, through which the waveguides are connected to a Vector Network Analyzer (VNA). Port 1 is excited while Port 2 serves as the receiver. The simulated and measured results as presented in Fig. \ref{figS7} show minimal reflection and good transmission at 2.35 GHz. The results show efficient coupling of the waves from the dielectric waveguide to the metallic waveguides.

\subsection*{The rotated Janus dipole}
\noindent
To enable efficient excitation of the Janus dipole, we employ a slightly different configuration in the microwave experiment, where the Janus dipole is rotated in the $xy$ plane, as shown in Fig. \ref{figS8}A. The helix is sandwiched by two waveguides with the gap distance of 50 mm. The incidence is a linearly polarized plane wave in the form of $\mathbf{E}_{\mathrm{inc}}=(-\cos \theta \hat{\mathbf{e}}_{x}-\sin \theta \hat{\mathbf{e}}_{y}) E_{0} e^{\left(\mathrm{i} k_{0} \sin \theta x-\mathrm{i} k_{0} \cos \theta y\right)}$, where $\theta$ is the incident angle between the wavevector $\mathbf{k}$ (on the $xy$ plane) and the $-y$ axis. The optimized Janus dipole can be achieved via tuning the rotation angle $\psi$ of the helix axis with respect to the $y$ axis on the $xy$ plane, i.e., tuning the match between the dipoles ($p_{x}$ and $m_{y}$) and the waveguide mode ($E_{x}$ and $B_{y}$) so that $\left|p_{x} E_{x}^{*}+m_{y} B_{y}^{*}\right| \rightarrow 0$, which gives $p_{x} /\left(m_{y} / c\right)=-c B_{y}^{*} / E_{x}^{*}=1.8 \mathrm{i}$, at incident angle $\theta=40$ degrees and rotation angle $\psi=35$ degrees, as shown in Fig. \ref{figS8}B.
\begin{figure}[h]
\centering
\includegraphics[scale=0.55]{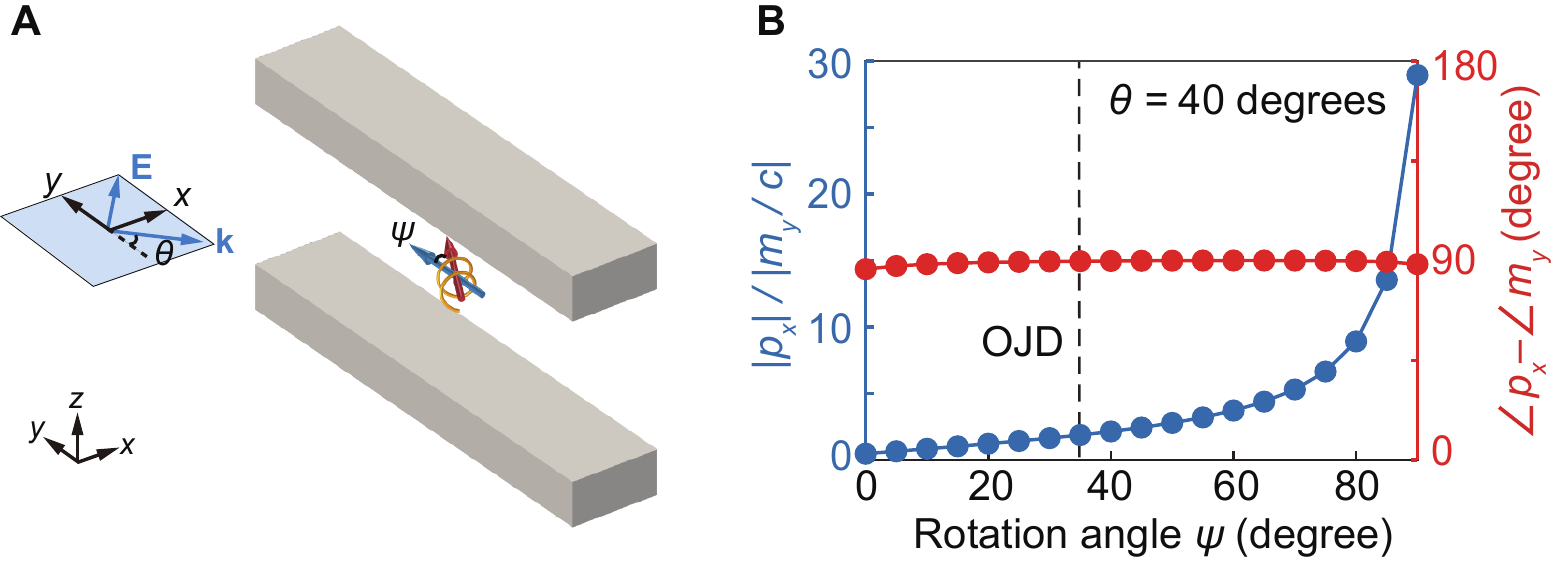}
\caption{\textbf{Janus dipole in the rotated helix}. (\textbf{A}) Schematic of the helix-dual-waveguide coupling configuration for demonstrating the directionality of the rotated Janus dipole. The blue and red arrows denote the $y$ direction and the center axis direction of the helix, respectively. (\textbf{B}) The relative amplitude and phase of the dipoles $p_{x}$ and $m_{y}$ as a function of the rotation angle $\psi$. The incident angle $\theta$ is set to be 40 degrees. } \label{figS8}
\end{figure}

\end{document}